\documentclass[conference]{IEEEtran}
\IEEEoverridecommandlockouts

\usepackage{amssymb} 
\usepackage{caption} 
\usepackage{booktabs}
\usepackage{makecell}
\usepackage{listings}
\usepackage[framemethod=TikZ]{mdframed} 
\usepackage{xcolor}
\usepackage{xspace}
\usepackage[inline,shortlabels]{enumitem}
\usepackage{multirow}
\usepackage{calc}
\usepackage{setspace} 
\usepackage{hyperref}
\hypersetup{
    colorlinks,
    linkcolor={black},
    citecolor={black},
    urlcolor={black}
}
\usepackage{balance} 
\usepackage[utf8x]{inputenc} 
\usepackage{newunicodechar}
\usepackage{tcolorbox}
\usepackage{soul}
\usepackage{bm}
\usepackage{wrapfig}
\usepackage{stfloats} 

\usepackage{ifthen}
\usepackage[normalem]{ulem} 

\newcommand{\eg}{\textit{e.g.\@}\xspace}

\newcommand{\ie}{\textit{i.e.\@}\xspace}

\newcommand{\diff}[2]{#2} 

\newcommand{\diffnote}[1]{} 

\newcommand{\sysraw}{Plottery}
\newcommand{\titletext}{\sysraw: A Code-Integrated GUI for \mpl}

\newcommand{\sys}{\textsc{\sysraw}\xspace}
\newcommand{\mpl}{Matplotlib\xspace}

\newcommand{\plotteryDotOrg}{\href{https://plottery.org}{\texttt{plottery.org}}\xspace}
\newcommand{\suppMaterialsRef}{\cite{PlotterySupplement}\xspace}

\newcommand{\studyone}{\textsc{Part~1}\xspace}
\newcommand{\studytwo}{\textsc{Part~2}\xspace}

\newcommand{\AI}{\emph{AI+Code}\xspace}
\newcommand{\DM}{\emph{GUI+Code}\xspace}
\newcommand{\Mixed}{\emph{Mixed}\xspace}

\renewcommand{\P}[1]{\ensuremath{\mathcal{P}\hspace{-1pt}\textit{#1}}}

\graphicspath{{images/}}

\newlength{\origcolumnsep}
\newlength{\origintextsep}

\DeclareUnicodeCharacter{9662}{\ensuremath{\blacktriangledown}} 
\newcommand{\btd}{\ensuremath{\blacktriangledown}\xspace}
\newunicodechar{▾}{\btd}

\newcommand{\citet}[1]{\cite{#1}} 
\newcommand{\cites}[1]{(\cite{#1})} 

\usepackage{inconsolata}  

\definecolor{CodeGrayColor}{RGB}{83, 83, 89}
\definecolor{CodeGrayColor}{RGB}{73, 73, 73}

\definecolor{CodeColor}{RGB}{29, 113, 171} 
\definecolor{GUIColor}{RGB}{208, 104, 11} 
\definecolor{AIColor}{RGB}{38, 138, 34} 

\newcommand{\code}[1]{\mbox{\small\texttt{\textcolor{CodeGrayColor}{#1}}}}

\newcommand{\mypara}[1]{\smallskip\noindent\textbf{\textit{#1.}}\xspace}

\newcommand{\onplotbutton}[1]{\tcbox[nobeforeafter,colback=darkgray,colupper=white,fontupper=\scriptsize\sffamily,boxrule=0mm,arc=0.3mm,left=0mm,right=0mm,top=0mm,bottom=0mm,height=3.5mm,tcbox raise=-0.8mm,valign=center]{#1}}

\definecolor{basicColor}{rgb}{0,0,0}
\definecolor{backgroundColor}{rgb}{.975,.975,.975}
\definecolor{frameColor}{rgb}{.85,.85,.85}
\definecolor{identifierColor}{rgb}{0.0,0.0,0.0}
\definecolor{keywordColor}{rgb}{.22,.49,.13}
\definecolor{stringColor}{rgb}{.67,.19,.16}
\definecolor{numberColor}{rgb}{.23,.53,.14}
\definecolor{operatorColor}{rgb}{.61,.18,.96}
\definecolor{commentColor}{rgb}{0.5,0.5,0.5}
\lstdefinestyle{notebookish}{
  columns=[l]fullflexible, 
  basicstyle=\linespread{1.0}\scriptsize\ttfamily\color{basicColor}, 
  breaklines=true, 
  keepspaces=true,
  breakatwhitespace=true,
  xleftmargin=0.3em,
  xrightmargin=0.3em,
  frame=none,
  framesep=0pt,
  belowskip=0.3em,
  aboveskip=0.3em,
  language=Python,
  showstringspaces=false,
  morekeywords={[1]True,False,None},
  otherkeywords={0,1,2,3,4,5,6,7,8,9,=,+,-,*,/}, 
  morekeywords={[3]0,1,2,3,4,5,6,7,8,9}, 
  morekeywords={[4]=,+,-,*,/}, 
  keywordstyle=\bfseries\color{keywordColor},
  keywordstyle=[2]\color{keywordColor}, 
  keywordstyle=[3]\color{numberColor}, 
  keywordstyle=[4]\bfseries\color{operatorColor},
  commentstyle=\color{commentColor},
  identifierstyle=\color{identifierColor},
  stringstyle=\color{stringColor},
  extendedchars=true,
  literate={∞}{{$\infty$}}1,
  escapeinside={<@}{@>}, 
}
\newenvironment{notebooklisting}
    {\begin{mdframed}[backgroundcolor=backgroundColor,
                     linecolor=frameColor,
                     linewidth=0.5pt,
                     innerleftmargin=0.2em,
                     innerrightmargin=0.2em,
                     innertopmargin=0em,
                     innerbottommargin=0em,
                     roundcorner=1.5pt
                     outertopmargin=0em,
                     ]
    }
    {
\end{mdframed}
    }

\DeclareRobustCommand*{\circled}[1]{\raisebox{-0.4ex}{\includegraphics[height=1.4\fontcharht\font`\B]{#1.png}\Description{#1}}\hspace{0cm}}

\newcommand{\Description}[1]{}

\newboolean{showedits}
\setboolean{showedits}{true} 
\ifthenelse{\boolean{showedits}}
{
	\newcommand{\del}[1]{\textcolor{red}{\sout{#1}}} 
}{
	\newcommand{\del}[1]{} 
	
}

\newboolean{showcomments}
\setboolean{showcomments}{true} 
\newcommand{\id}[1]{$-$Id: scgPaper.tex 32478 2010-04-29 09:11:32Z oscar $-$}

\ifthenelse{\boolean{showcomments}}
{\newcommand{\nbc}[3]{
 {\colorbox{#3}{\bfseries\sffamily\scriptsize\textcolor{white}{#1}}}~
 {\textcolor{#3}{\sf\small$\blacktriangleright$\textit{#2}$\blacktriangleleft$}}}
 }
{\newcommand{\nbc}[3]{}
 \renewcommand{\del}[1]{} 
 }

\definecolor{ibcolor}{rgb}{0.4,0.6,0.2}

\definecolor{mycolor}{rgb}{0.1,0.4,0.1}
\definecolor{lightblue}{rgb}{0.9,0.9,1.0}

\begin{document}

\title{\titletext}

\title{\titletext%
}
\IEEEpubid{\begin{minipage}{\columnwidth}\vspace{1em}©2026 IEEE. This is the authors' accepted version of the paper. The revised final version is in VL/HCC 2026.\end{minipage} \hspace{\columnsep}\makebox[\columnwidth]{ }}


\newcommand{\markUCSD}{\textsuperscript{\raisebox{-1.2pt}{*}}}
\newcommand{\markCornell}{\textsuperscript{\raisebox{-1pt}{$\dagger$}}}
\author{\IEEEauthorblockN{Brian Hempel\markUCSD,
Ruanqianqian (Lisa) Huang\markCornell\markUCSD,
Devamardeep Hayatpur\markUCSD,
Sorin Lerner\markCornell,
Haijun Xia\markUCSD}
\begin{minipage}{0.49\textwidth}
\vspace{0.75pt}
\centering
{\markUCSD}\{bhempel,
dshayatpur,
haijunxia\}@ucsd.edu\\
University of California San Diego, La Jolla, CA, USA
\end{minipage}
\hspace{0.5pt}
\begin{minipage}{0.34\textwidth}
\vspace{0.75pt}
\centering
{\markCornell}\{lisahuang,
sorin.lerner\}@cornell.edu\\
Cornell University,
Ithaca, NY, USA
\end{minipage}
}

\IEEEaftertitletext{
\includegraphics[width=1.003\textwidth]{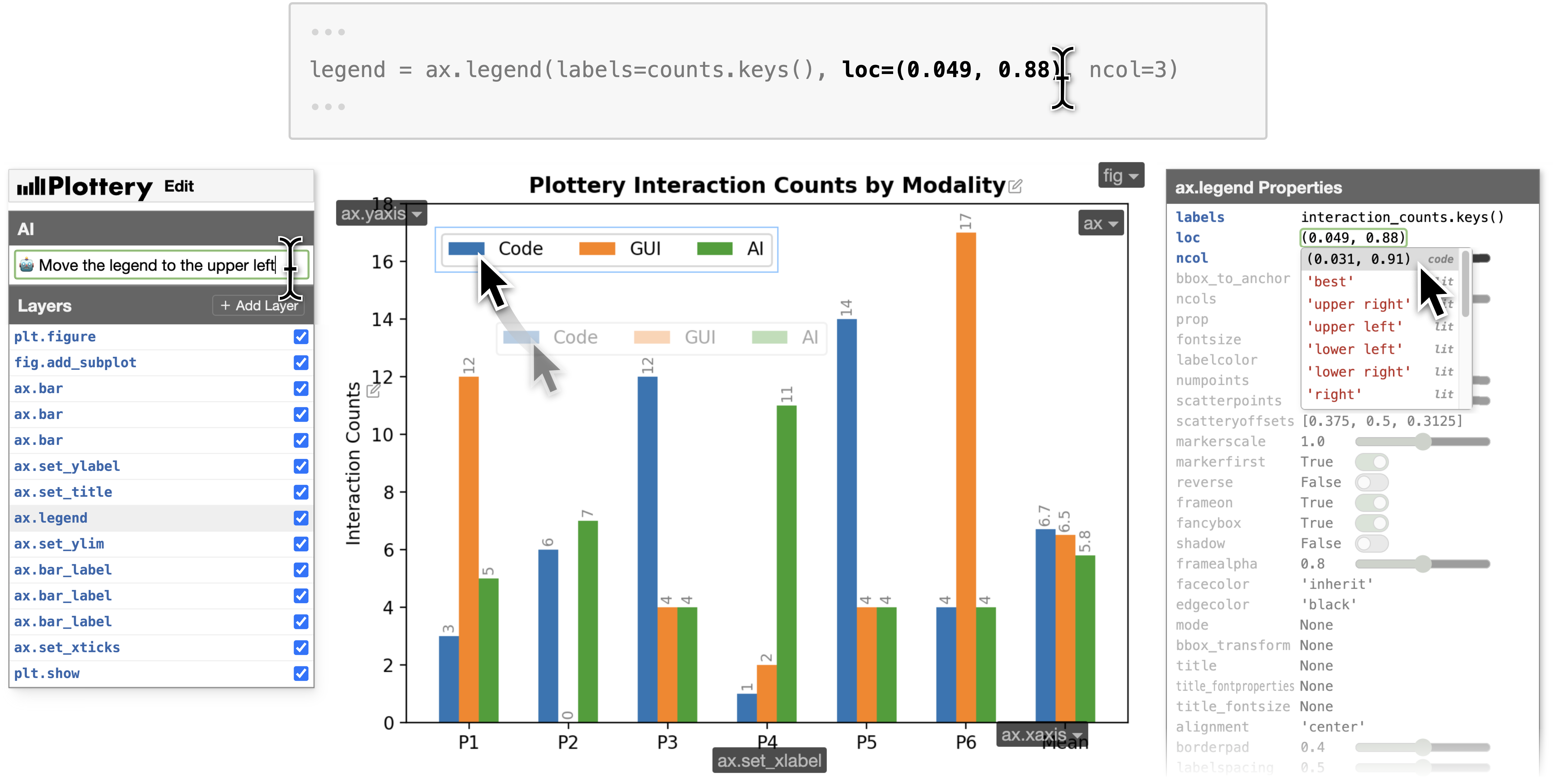}
\captionof{figure}{\sys imitates graphical editors, offering live GUI editing of \mpl code in an ordinary Jupyter notebook. \\[0.5em]}
\label{fig:teaser}
}


\maketitle

\begin{abstract}
Scientists have many tools to author plots: graphical direct manipulation apps like a spreadsheet's chart builder,  text-based programming tools like \mpl and ggplot2, or AI chat interfaces.
Textual code is the most capable of these, but writing code, even with AI assistance, is a potentially poor fit for graphical chart design.
We present \sys, a Jupyter notebook extension that leverages API type annotations to add graphical direct manipulation to the popular \mpl plotting library.
User interactions in \sys's in-situ graphical interface are immediately reified in the user's code, so they may seamlessly mix-and-match between GUI, code, and AI interactions.
In a user study, participants seamlessly alternated modalities, with the GUI used for straightforward visual changes, code for quick edits or duplicating code, and the AI for broad support.
\sys showcases graphical interaction that retains code's benefits.
\end{abstract}


\section{Introduction}\label{sec:introduction}

With the ubiquity of scientific data plotting, plot authors have many ways to create their plots.
They might reach for a \emph{graphical user interface (GUI)}, such as the charting features in a spreadsheet.
They might write \emph{code}, using a library such as Matplotlib, ggplot, or Vega-lite.
Or, they might upload their data to a chatbot to build their plot with \emph{AI} through natural language. Each approach has pros and cons.

GUIs are perhaps the most popular way to create plots, from commercial tools such as spreadsheet applications, Tableau, and Power BI to academic tools with innovative interfaces like Charticulator~\cite{Charticulator}, Data Illustrator~\cite{DataIllustrator}, Lyra~\cite{Lyra}, and others~\cites{iVisDesigner, InfoNice, DataDrivenGuides, VisComposer}.
In GUIs, the available actions are discoverable in the interface, direct manipulation lets users precisely modify the visual design with their mouse cursor, and changes are displayed immediately for rapid iteration~\cite{DirectManipulation}.
Graphical interfaces naturally fit graphical plots.

Nevertheless, code-based plotting libraries like Matplotlib~\cite{matplotlib}, ggplot2~\cite{ggplot2}, or Vega-Lite~\cite{VegaLite} remain common.
Although foregoing the pleasant experience of direct manipulation~\cite{DirectManipulationCogSci}, text-based tools offer at least three advantages:
code can be written \emph{in-situ} within a data analysis framework, such as a Jupyter notebook, enabling rapid iteration between analysis and visualization~\cites{HowDataAnalystsUseAVisualizationGrammarInPractice,grammel2013visSurvey};
code offers near-unlimited \emph{expressivity}, allowing authors to customize the plot as desired~\cites{grammel2013visSurvey,chopra2023whyCodeForVis,archiepomchi2024matplotlibVersatile};
and code can \emph{\mbox{automate}} plot production, for reproducibility~\cites{fivePillarsOfComputationalReproducibility,chopra2023whyCodeForVis} or for drawing many similar plots\diff{}{~\cite{chopra2023whyCodeForVis}}. Still, writing code can be tedious,
for example \mpl, the target of this work, is disliked by some of its users~\cites{jachymb2022matplotlibBadLibrary,deleted2021despisesMatplotlib,inconspicuousmale2022matplotlibSucks} for its ``garbage'' usability~\cite{NFeruch2024matplotlibGarbageLibrary},
a problem exacerbated because,
even in the AI era, using such code-based tools sometimes still requires the laborious process of looking up documentation, copy-pasting a code example, and modifying it to fit one's scenario.

To avoiding the learning curve of complicated code APIs, plot authors may instead create their plots with natural language~\cites{TowardsNaturalLanguageInterfacesForDataVisualizationASurvey,TheWhyAndTheHowASurveyOnNaturalLanguageInteractionInVisualization}.
Current leading AI tools can draw charts based on data uploaded to the chat interface (\eg~\cite{aiChartingTutorial}) or from within a spreadsheet integration~\cites{aiChartsInExcel,aiChartsInGoogleSheets}.
Prompt-based AI interfaces offer benefits: users can describe complex charts using imprecise language and achieve their goals with little expertise in the plotting library. 
However, checking if AI-generated code matches the user's intent can be difficult~\cites{ValidatingAiGeneratedCodeWithLiveProgramming,mozannar2024reading}, and a GUI may be better for fine-grained graphical control and exploring design tweaks, like repositioning a legend on a plot or picking specific colors~\cites{DynaVis,DirectManipulationCogSci}.

 \textbf{Our goal is to augment plot code editing with a seamlessly integrated GUI, \emph{without} losing any of code's benefits}, 
namely, code's operation \textit{in-situ} in the data science workflow, its \textit{expressivity}, and its ability to \textit{automate} plot creation.
Although there are capable code-generating plotting GUIs~\cites{Lyra,mage,Blace,CodeMend,Pylustrator},
in our view there is still no interface that integrates GUI and code seamlessly, with \textit{both} as powerful first-class interaction modes.
To demonstrate our ideal vision of a code-integrated plotting GUI,
we target a popular but difficult API: the \mpl library~\cite{matplotlib} within Jupyter notebooks~\cite{Jupyter}. Although \mpl is the most popular Python plotting library (as measured by Google search results), it is in dire need of better tooling
because of its unintuitive API~\cites{jachymb2022matplotlibBadLibrary,NFeruch2024matplotlibGarbageLibrary}, which earns comments from users such as \emph{``Every time I need to use mpl for a project I die a little inside''}~\cite{deleted2021despisesMatplotlib} and
\emph{``Matplotlib sucks and we all know it. It's just the only library that can do everything we need it to and has documentation detailing exactly how to do it''}~\cite{inconspicuousmale2022matplotlibSucks}.
To help, we implement a notebook extension called \sys.
As the craft of pottery mixes clay and water to sculpt a concrete object,
in \sys users mix GUI and (AI-assisted) code interactions to fluidly sculpt \mpl plots.

\mypara{Design Goals}
To simultaneously be as ``GUI-like'' and ``code-like'' as possible, \sys aims to provide:

\begin{enumerate}[leftmargin=*]
	\item \textbf{Familiarity} by imitating a traditional graphics editor.
	\item \textbf{Fluid iteration} through direct manipulation~\cites{DirectManipulation,DirectManipulationCogSci}.
	\item \textbf{Predictability} in how GUI actions become code.
	\item \textbf{Compatibility} with code's benefits: in-situ operation, expressive authoring, and automation. Specifically, it should work in the notebook, support much of the \mpl API, and support loops and functions.
	\item \textbf{Fluid mode switching}: users should be able to fluidly bounce between GUI, code, and AI interactions without the friction of switching costs~\cites{mage,WhyProgrammingByDemonstrationSystemsFail}.
\end{enumerate}

\noindent
To meet these goals, \sys adopts two simple but novel design commitments:

\begin{enumerate}[leftmargin=*]
  \item \textbf{Calls = Layers.} \sys treats individual \mpl API calls as \emph{layers} in a traditional graphics editor, organizing them into a layers panel for selection, toggling, and reordering (left in \autoref{fig:teaser}).
  \item \textbf{Arguments = Properties.} The arguments in each individual call are analogs to \emph{item properties} in a traditional editor. \sys uses \mpl API type annotations to present possible call arguments in a properties editor, with type-appropriate UI widgets for each property (right in \autoref{fig:teaser}). This type-driven approach allows \sys to support a meaningful portion of the \mpl API.
\end{enumerate}

\noindent
To show how \sys implements design goals 1-4, after discussing related work (\autoref{sec:related}) we demonstrate \sys in an example walkthrough (\autoref{sec:overview-example}) followed by a description of the implementation of its key features (\autoref{sec:implementation}). For design goal~5, fluid mode switching, we investigate user interaction with GUI, code, and AI in \sys through an exploratory user study (\autoref{sec:eval}). This study provides evidence not only for \sys's usability and seamless multimodality, but also provides insights on the complementarity of GUI, code, and AI.
In particular, we find
(1) all modalities were used, users leaned most on AI but individual preferences varied;
(2) each modality offers its own complementary strengths—GUI suits straightforward visual changes; AI helps with getting started and when unsure how to make a change; and code is used for quick, small edits and for copy-paste-modify duplication;
(3) as we hoped, the interaction of all three modalities was \emph{seamless}, study participants switched between modes after only about 3 interactions on average.
We discuss these results (\autoref{sec:discussion}) and conclude (\autoref{sec:conclusion}).

\section{Related Work}\label{sec:related}


Below we survey graphical tools for bespoke visualizations, natural language for plotting, and multimodal plot authoring.




\mypara{GUIs for Custom Visualizations} 
GUI-based systems, dating back to at least Gold~\cite{Gold} and SageBrush~\cite{SageBrush}, use a vector graphics editor and bind properties of primitive shapes to data
 to produce customized data-bound visualizations.
While some systems focus on glyph (mark) customization~\cites{DataInk, InfoNice}, others target better support for user interaction and data use.
To support bulk edits of visualizations, some systems represent repeated elements as groups on the canvas~\cites{Lyra, iVisDesigner, Charticulator}, while others capture user actions as step-by-step (non-textual) programs with loops to enable reuse~\cite{DrawingDynamicVisualizations}.
These systems also enable data use in various ways, from letting users design visuals before connecting with data~\cites{DataIllustrator, StructGraphics}, abstracting the spatial representation of repeated data into hidden layout guides~\cite{DataDrivenGuides}, enabling data binding specifications on a separate node-and-wires programming canvas (that could be further used for data transformation)~\cite{VisComposer}, using a spreadsheet paradigm~\cite{Cuscus}, or inferring visualizations and the relevant data transformations from user examples of data bindings~\cite{Falx}.
We provide expressive authoring not by innovative interactions as in the above systems, but by following the expressive \mpl API so \sys can operate in-situ in the user's Jupyter notebook, which streamlines automated regeneration of plots by avoiding export-import to a separate application.

\mypara{Natural Language for Visualizations}
Users may not know how to create their desired plot with a complex GUI---a difficulty dubbed the \emph{gulf of execution}~\cite{DirectManipulationCogSci}--so it would be helpful if they could describe their visualization in terms they understand.
Consequently, natural language has been applied, successfully, to visualization even before the recent rise of large language models (LLMs)~\cites{TowardsNaturalLanguageInterfacesForDataVisualizationASurvey, TheWhyAndTheHowASurveyOnNaturalLanguageInteractionInVisualization}.
Nevertheless, newer LLM-based systems can outperform classical NLP for plotting~\cite{ChartGPT},
and flagship LLMs are skilled at generating \mpl code~\cites{CHAT2VIS,LIDA}, with one user quipping \emph{``matplotlib is actually really awful and it's hilarious that we had to invent literal AGI for me to start using it''}~\cite{kache2024matplotlibAwful}.
\sys makes no innovation here, but includes an AI prompt box to allow natural language authoring alongside GUI and code editing.

\mypara{Natural Language and GUI}
Although natural language can help build up a visualization without learning a GUI,
Vaithilingam~et~al.~\cite{DynaVis} hypothesize that a GUI is better for \emph{tweaking} plots, such as picking specific colors. 
Combining natural language with a GUI may provide the complementary strengths of each~\cites{VisTalk, DynaVis}.
VisTalk~\cite{VisTalk} uses a non-LLM approach to transform keywords in the user's input (\eg ``red'') into clickable regions that open a widget (\eg a color picker).
Augmenting this approach, DynaVis~\cite{DynaVis} lets users prompt an LLM to create or change their plot (represented internally as Vega-Lite~\cite{VegaLite}) and simultaneously generates a \emph{dynamic widget} to control relevant plot properties related to the prompt (\eg number inputs for editing the Y-axis limits). 
Users can also specifically ask the LLM to make a GUI widget.
Although not targeting plotting, BISCUIT~\cite{BISCUIT} similarly generates transient widgets from prompts; widgets steer AI generation of code in the following notebook cell.
LLM-generated widgets can randomly vary from prompt to prompt.
Because our goal is to innovate on code-integrated GUIs and not AI interaction, \sys generates its UI deterministically based on \mpl API type annotations rather than requiring users to prompt AI to create a GUI.

\mypara{Code and GUI}
Some Python-focused plugins~\cites{bamboolib, Mito, Dtale} offer GUI-based chart building \emph{in-situ} within the user's data analysis environment (\eg a Jupyter notebook), generating code from the GUI.
However, code generation is one-way: text edits to the code are not reflected back into the GUI. B2~\cite{B2}
can reflect code edits back into the GUI but its GUI actions, like the above, only generate code rather than also edit it.


Three prior systems allow \diff{}{full} round-trip editing between code and the GUI to truly mix the modalities~\cites{Pylustrator, CodeMend, mage}.
Pylustrator~\cite{Pylustrator} takes a base plot built from \mpl code as input, allows users to further modify the plot from the GUI, and saves GUI edits to \mpl code appended to the original code. These edits are limited to moving and tweaking items---Pylustrator does not allow plotting new data within the GUI.
Round-trip editing is possible but tedious: a user can edit the generated code then reopen it in Pylustrator for further GUI-based changes.
CodeMend~\cite{CodeMend} combines natural language, code, and a GUI property editor, taking a natural language query from the user to highlight the most relevant calls/properties in the GUI for the user to interact with to update their code.
CodeMend's GUI property editor is rather full-featured, but its overall interface was designed for demonstrating pre-LLM natural language input so it lacks the smooth editing of a traditional graphics editor that we seek.
For that goal, the closest to our system is mage~\cite{mage} which mixes live direct manipulation with code editing in Jupyter notebooks. Although its GUI interaction better imitates the fluid editing of a traditional graphical editor than CodeMend, mage's small plotting GUI only supports a limited subset of Vega-lite, rather than aiming for highly-customizable Python-integrated authoring.

Our goal in \sys is to demonstrate an ideal code-integrated GUI for plot authoring.
\sys combines Pylustrator's direct manipulation on the plot, CodeMend's more complete treatment of the \mpl API, and mage's fluidity in-situ in the notebook. To this, \sys adds support for functions and loops, and presents all this like a traditional graphical editor but produces always-editable Python code.

\section{Plottery}\label{sec:overview-example}

\newcommand{\minihist}[1]{\raisebox{-0.365em}{\includegraphics[height=1.4em]{overview_ex_mini_hist_#1}}}

\sys is a Jupyter notebook extension
that augments \mpl code cells with a GUI,
aiming to empower users to smoothly construct non-trivial plots with more pleasure and less fumbling than is usual with \mpl.
%
%
%
%
%
To demonstrate \sys, we will build a bare bar chart, like \minihist{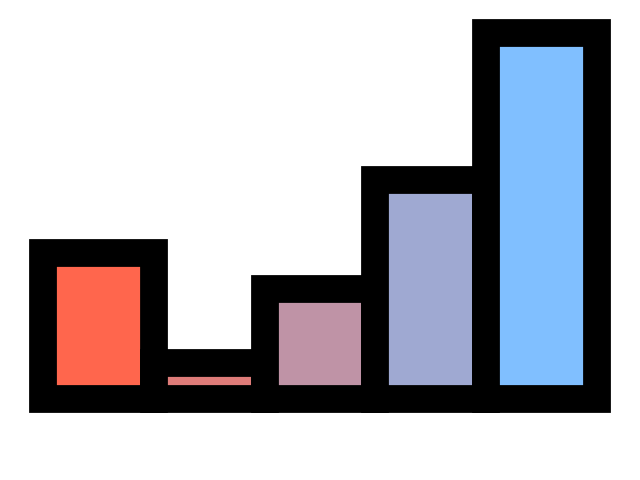}\Description{a mini histogram colored with a gradient from left to right}, suitable for inclusion inline in text for, \eg{}, reporting Likert score counts. We will turn this into a function to reuse the plot with different data to produce three plots: \minihist{no_dot_1} \minihist{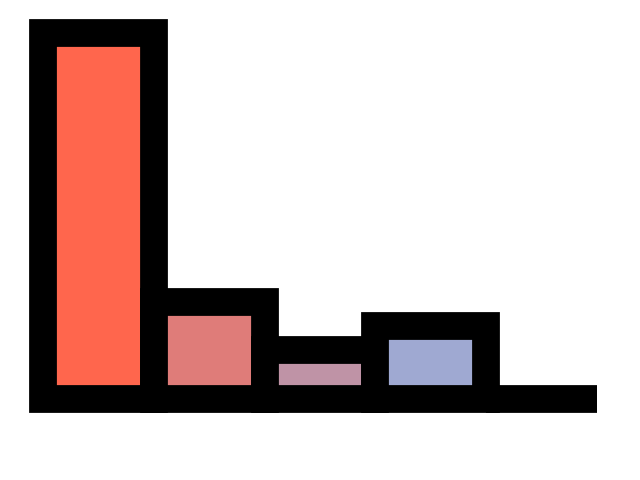}\minihist{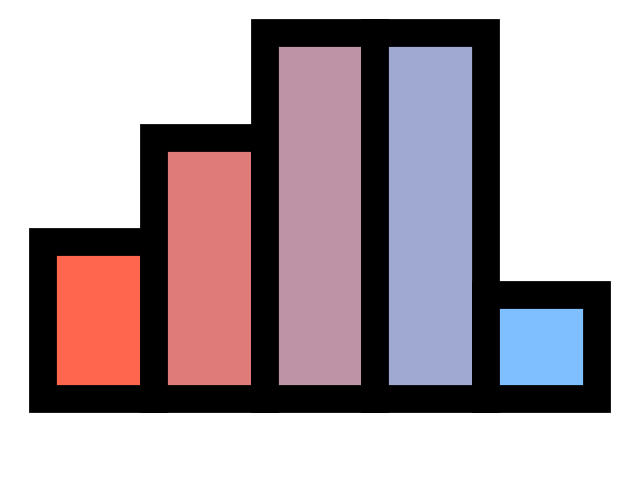}\Description{3 mini histograms}. Then we will add a dot for the mean, like  \minihist{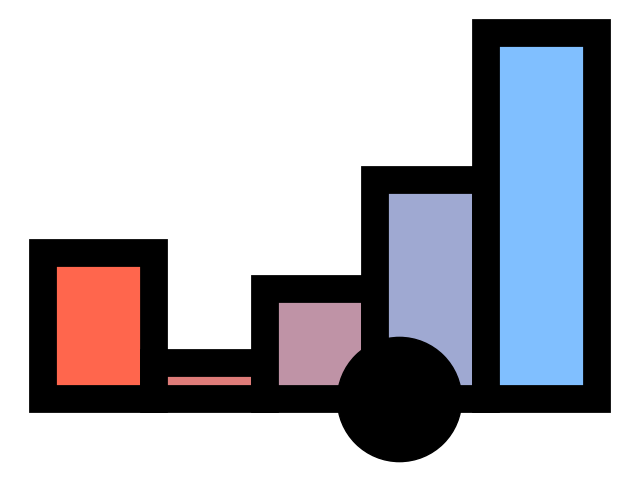} \minihist{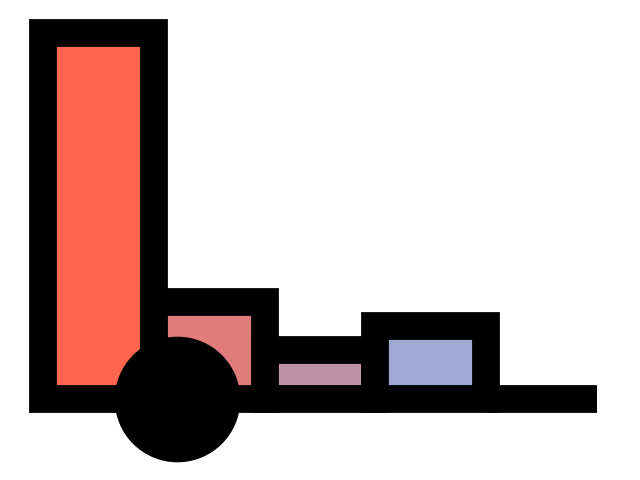} \minihist{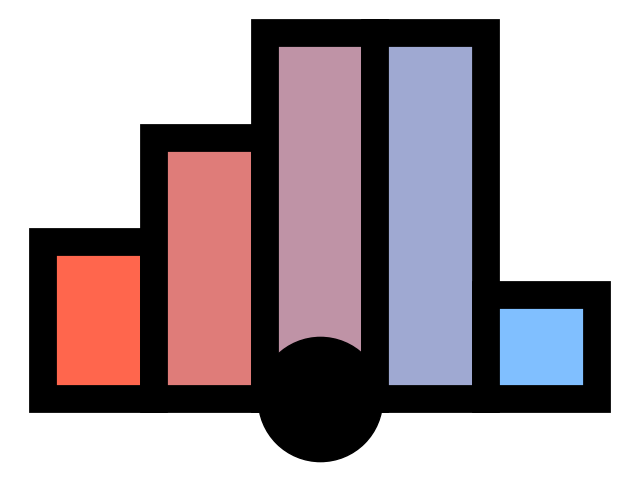}\Description{3 mini histograms each with a dot on the baseline}, to demonstrate how \sys works even after abstracting the plot into a function.
%
We will use all three interaction modalities \sys supports: GUI, code, and AI. \emph{Our anonymous supplement at \plotteryDotOrg~\suppMaterialsRef includes a narrated five minute video figure recording of this walkthrough.
}
\begin{figure}[b]
\centering
\begin{minipage}{0.6\columnwidth}
\begin{notebooklisting}\begin{lstlisting}
counts = [4, 1, 3, 6, 10]
\end{lstlisting}\end{notebooklisting}
\vspace{-0.8em}
\begin{notebooklisting}\begin{lstlisting}
fig = plt.figure(layout='tight')
ax = fig.add_subplot(1, 1, 1)

plt.show()
\end{lstlisting}\end{notebooklisting}
\vspace{0.5em}
\end{minipage}
\includegraphics[width=\columnwidth]{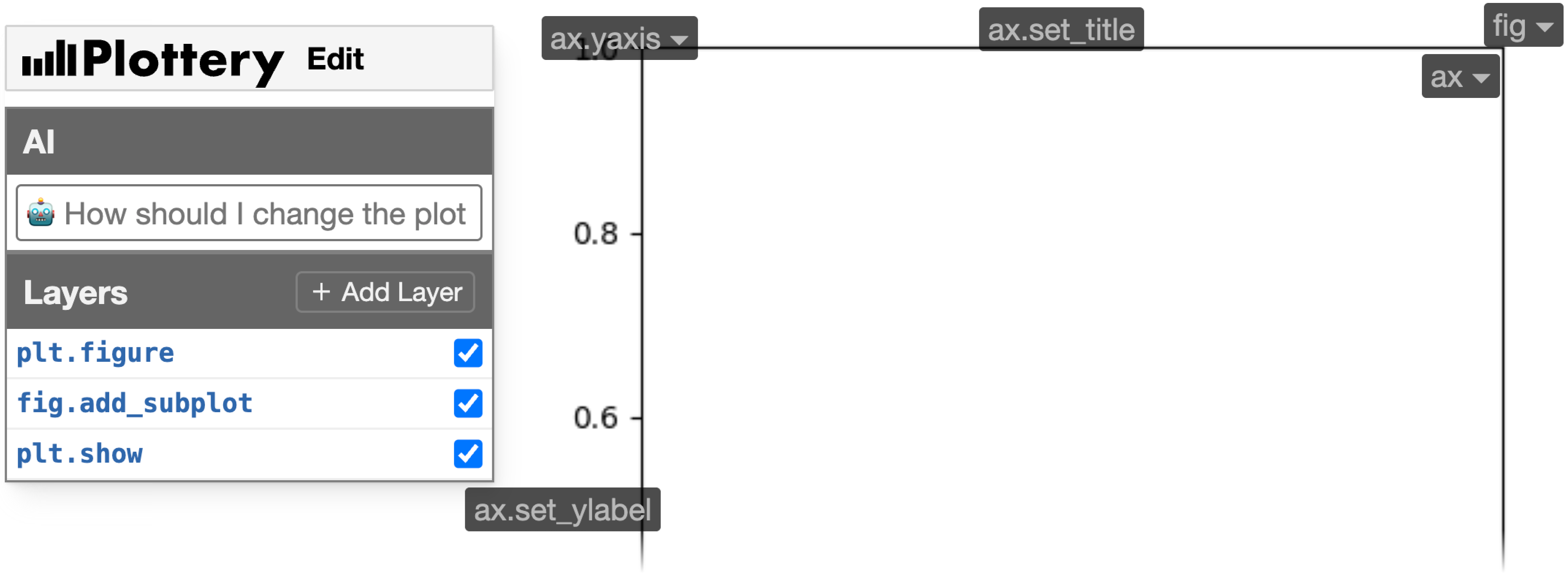}
\caption{Example data, starter code, and the initial \sys UI next to a new plot with an empty axes box. The left sidebar has an AI prompt box and a Layers panel listing \mpl calls in the code. Buttons on the plot allow adding new code.}
\label{fig:ex_start}
\vspace{-0.4em} 
\end{figure}

\begin{figure*}
    \centering
    \includegraphics[width=\linewidth]{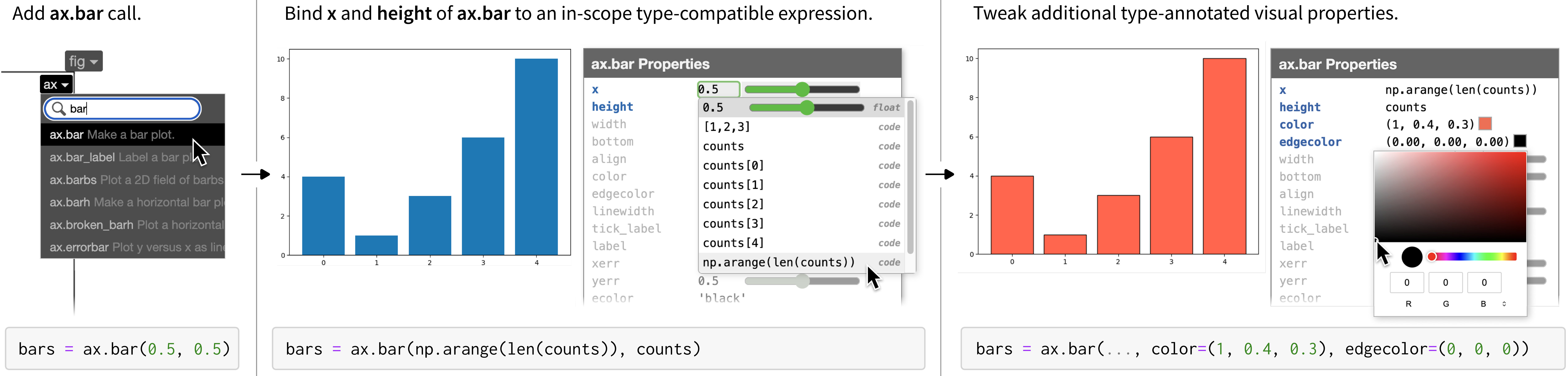}
    \caption{Adding an \code{ax.bar} call from the \textnormal{\onplotbutton{ax \btd}} dropdown, and binding and styling the bars with the Properties panel.}
    \label{fig:ex_making_bars}
\end{figure*}

\mypara{Getting Started}
We fill in a code cell with example data in a \code{counts} variable (first cell in \autoref{fig:ex_start}). \sys provides a ``New Plot'' button in the Jupyter interface (not shown), we click it which adds starter code in a new cell (second cell in \autoref{fig:ex_start}). This code creates a figure and adds one \emph{axes}, \ie cartesian coordinate box, to it. \sys's UI appears when we run the cell (\autoref{fig:ex_start} bottom).

\mypara{Layers Panel}
\sys's left sidebar has an AI prompt box and a \emph{Layers panel} that lists the code's \mpl calls (\autoref{fig:ex_start} bottom left). In the Layers panel, clicking a layer selects the call to edit its properties, drag-and-drop reorders calls, and clicking the visibility checkbox comments or uncomments the associated line of code. These latter two features are like the R data analysis tool Unravel~\cite{Unravel}, but for plotting.

\mypara{On-Plot Buttons}
When the mouse cursor is over the plot, \emph{on-plot buttons} appear for adding new calls to our cell (\autoref{fig:ex_start} bottom right). For example the \onplotbutton{ax.set\_title} button would add \code{ax.set\_title('My Plot')}.
The buttons with \btd open a menu of possible function calls, with a search box to filter the many items (\autoref{fig:ex_making_bars} left). We click \onplotbutton{ax \btd} and choose \code{ax.bar} to add a bar plot, which adds the code \code{bars = ax.bar(0.5, 0.5)}.


\mypara{Properties Panel}
The updated code cell is automatically re-run, the new \code{ax.bar} call is selected in the left sidebar (not shown),
and a \emph{Properties panel} appears on the right listing both given arguments and greyed-out possible arguments for the selected call (\eg \autoref{fig:teaser} right).
The traditional way to work with \mpl is to search online for a function's documentation, find a desired argument, write it in our code, and rerun. The Properties panel makes this process instant. Clicking an argument toggles it on or off, with the plot live updating to show the effect. This enables rapid exploration.

We see our bar's \code{x} and \code{height} arguments are literal numbers \code{0.5} and \code{0.5}.
We want to bind these arguments to our \code{counts} data.
We click on the number \code{0.5}, which lets us text-edit its value but also opens a dropdown of suggested values (\autoref{fig:ex_making_bars} middle).
The suggestions are based on the type annotation for the \code{x} argument and the variables in scope that match that type, including depth 1 indexing into those variables (\eg \code{counts[0]}) and common expressions involving those variables (\eg \code{np.arange(}\hspace{0pt}\code{len(counts))}).
The plot display live updates as we hover over each, previewing the result.
We choose \code{np.arange(}\hspace{0pt}\code{len(counts))}, which is the numbers \code{[0, 1, 2, 3, 4]}, as the \code{x} positions for our bars. For \code{height}, we choose \code{counts}, resulting in five bars (\autoref{fig:ex_making_bars} middle).

\begin{figure}[b]
\centering
\includegraphics[width=\columnwidth]{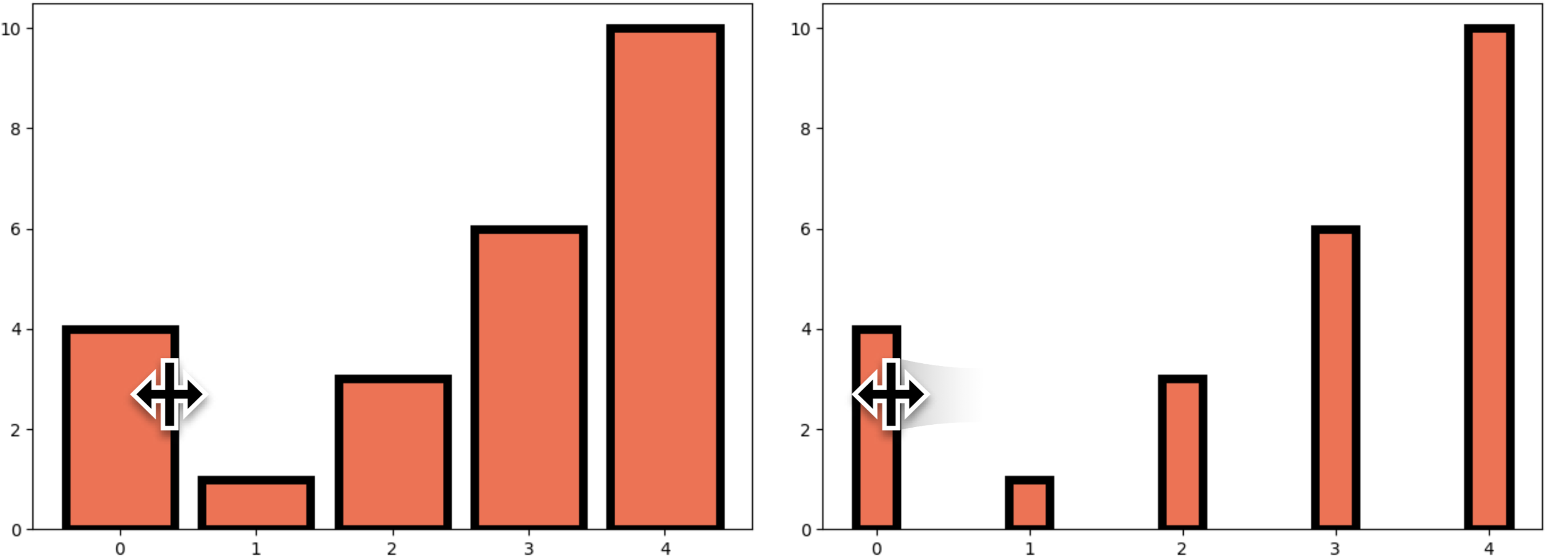}
  \caption{Some spatial properties can be manipulated on the plot.}
  \label{fig:ex_resizing_width}
\Description{
Three images of the bar plot with original width, skinny, and fat flush bars respectively. A resize cursor on the edge of the first bar is resizing the bars.
}
\vspace{-0.3em} 
\end{figure}

Styling is similarly quick. Color arguments have a color picker widget—we use these to set the bar \code{color} to a shade of orange and \code{edgecolor} to black (\autoref{fig:ex_making_bars} right). Numeric arguments can be controlled with a slider \raisebox{-0.2em}{\includegraphics[height=1em]{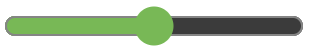}\Description{horizontal slider}}. We slide the \code{linewidth} to \code{5} to thicken their borders.
Each of these changes is instantly realized in the code and on the plot.

\mypara{On-Plot Direct Manipulation}
We want our bars to touch each other. \sys allows certain parameters to be modified by direct manipulation on the plot itself. We grab the left or right edge of any bar and drag horizontally (\autoref{fig:ex_resizing_width}). This adds a \code{width} argument to our \code{bar} call and modifies \code{width} to match our mouse movements live—we release the mouse when the bars touch each other at \code{width=1}. Hitting \code{1} exactly is possible because \sys rounds numbers to two significant figures during drags to imitate ``snap to grid'' in graphics editors.

%



\mypara{Using AI}
Recall our goal is to produce miniature histograms like \minihist{1} \minihist{2} \minihist{3}\Description{3 mini histograms each with a dot on the baseline}.
We now want to remove the border and numbers around our plot. Doing this through the GUI is possible but requires a few steps which we might not know.
In the left sidebar's \emph{AI panel}, we prompt ``Remove the borders around the plot'' and then ``Remove the numbers along the axes''.
After each, \sys replaces the code cell with the LLM's response, highlights changed lines, and runs the cell to show the updated plot.
 We also ask the AI to color our bars with a gradient like \minihist{no_dot_3}\Description{mini histogram with bars colored from orange to blue}.

%
%

\begin{figure}[t]
\centering
\begin{minipage}{0.6\columnwidth}
\begin{notebooklisting}\begin{lstlisting}
 counts_lists = [
     [4,  1, 3, 6, 10],
     [15, 4, 2, 3, 0],
     [3,  5, 7, 7, 2],
 ]
\end{lstlisting}\end{notebooklisting}
\vspace{-0.8em}
\begin{notebooklisting}\begin{lstlisting}
def plot_counts(counts):
   ...
   plt.show()

for counts in counts_lists:
   plot_counts(counts)
\end{lstlisting}\end{notebooklisting}
\vspace{0.2em}
\includegraphics[width=\textwidth]{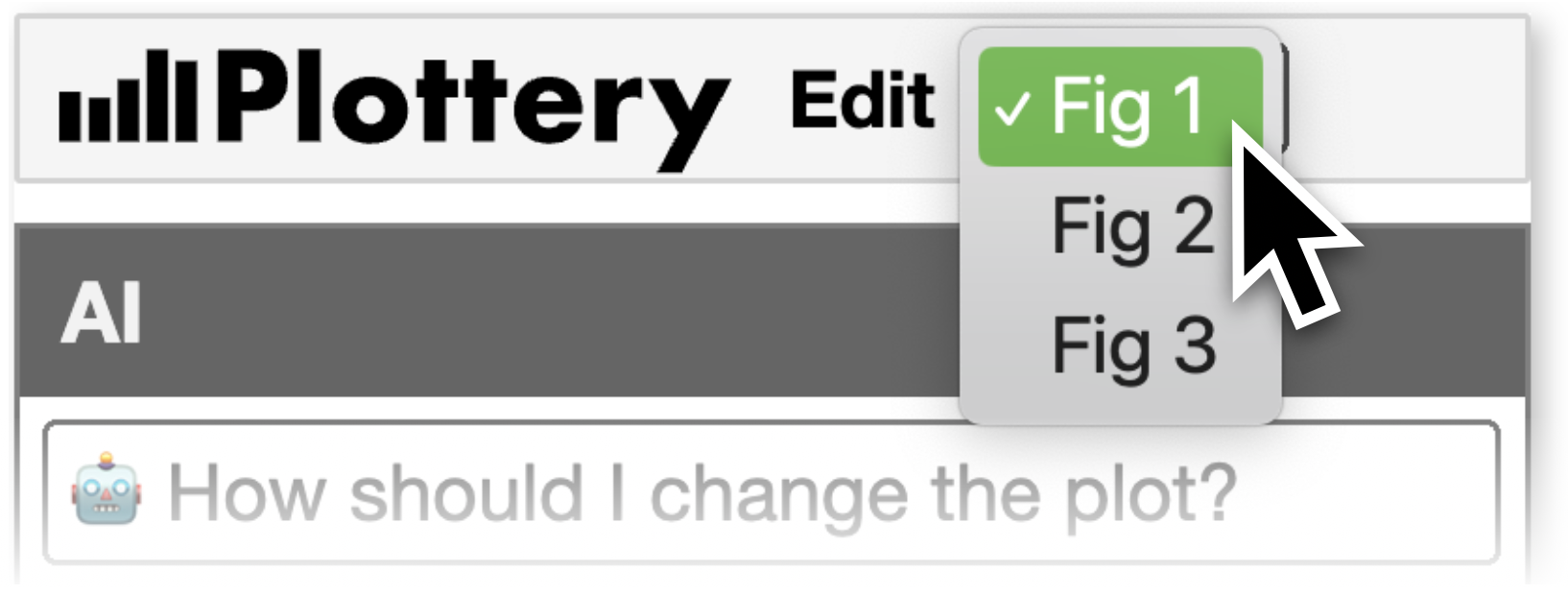}
\end{minipage}
\caption{\sys lets us pick which of multiple plots to view.}
\label{fig:ex_automation}
\Description{
The left sidebar header now includes a dropdown with Fig 1 Fig 2 Fig 3.
}
\end{figure}

\mypara{Automation}
A key reason to plot using code is for reuse: we want to make many similar bar charts.
\autoref{fig:ex_automation} shows how we set this up in our notebook: we add data for the three plots in a \code{counts\_lists} variable, wrap all of our existing plotting code in a \code{plot\_counts} function, and then call that function in a for-loop to plot each data series. AI could help us write this code, but here, it is simple enough to do manually.
Our code now produces three plots. \sys only shows one but provides a dropdown to switch between them (\autoref{fig:ex_automation} bottom).

\begin{figure}[h]
\centering
\includegraphics[width=0.9\columnwidth]{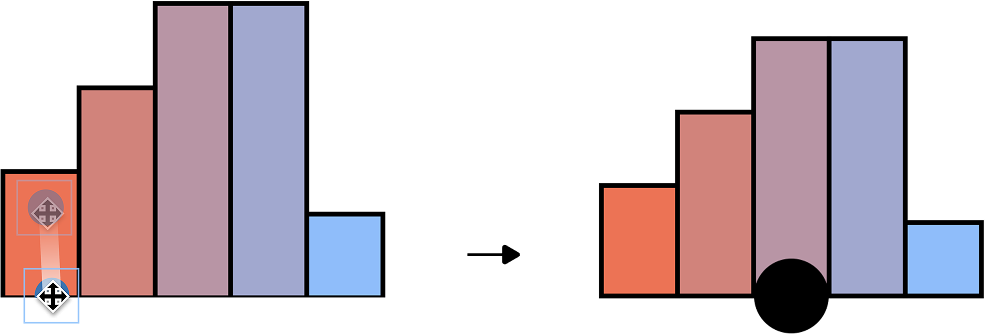}
\begin{minipage}[c]{0.9\columnwidth}
\vspace{0.4em}
\begin{notebooklisting}\begin{lstlisting}
def plot_counts(counts):
   mean = ...
   ...
   paths = ax.scatter(mean, 0, s=6000, c=(0.00, 0.00, 0.00))

   ax.set_ylim(bottom=-0.15*max(counts))

   plt.show()
\end{lstlisting}\end{notebooklisting}
\end{minipage}

  \caption{Final graphical changes and code.}
  \label{fig:ex_final}
\Description{
Two versions of the gradient-color bars.
In the first, a move cursor drags a blue dot down to the baseline.
In the second, the dot is larger, black, and positioned on the middle of the baseline.
}
  
\end{figure}

GUI edits work even when the plotting code is inside a function or loop.
Recall we want to add a dot to indicate the mean, like~\minihist{3}\Description{mini histogram with a dot on the baseline}.
After manually writing the computation to compute the mean, we click~\onplotbutton{ax \btd} on the plot to add \code{ax.scatter} (a scatterplot), which will, by default, draw a single circular dot.
\sys inserts code relative to the code location of \code{plt.show()}, so \code{ax.scatter} is inserted \emph{inside} our function.
Direct manipulation continues to work. The dot starts out at \code{(0.5, 0.5)}, which we drag down to a \code{y} of \code{0} (\autoref{fig:ex_final} left) and then use the Properties panel to set the remaining arguments: we choose \code{mean} for \code{x}, set the color to black, and drag the \code{size} slider until the dot is large. The bottom half of the dot is hidden because it extends outside the axes frame, so we add an \code{ax.set\_ylim} call and set the \code{bottom} of the axes to a negative value instead of 0. \autoref{fig:ex_final} shows the final design (right) and the code produced by these final changes. We ``Save'' from the menu to export PNG, and our plots are~done:~\minihist{1}~\minihist{2}~\minihist{3}\Description{3 mini histograms each with a dot on the baseline}!

\mypara{Recap}
Above we showed \sys's multimodal plot authoring: the \emph{Layers panel} enables code reordering and selecting calls for editing, the plot area features \emph{on-plot buttons} like \onplotbutton{ax \btd} for adding items, certain spatial properties may be tweaked by \emph{on-plot direct manipulation}, a call's arguments are editable via the \emph{Properties panel} with widgets and suggestions for each argument, the GUI supports functions or loops that draw multiple plots, the \emph{AI Panel} offers edits by prompting, and \emph{manual code edits} are always available at any time.

\section{Implementation}\label{sec:implementation}

We briefly explain some of \sys's technical details.
Hurried readers should only read the Properties panel section.

\mypara{Notebook Extension}
\sys consists of a Jupyter Notebook extension, a Python backend, and a Typescript front-end that renders the GUI. The notebook extension installs a Jupyter hook to silently rewrite code before it is sent to the Python backend, replacing calls to \code{plt.show()} with 
a \sys-specific call that tells the Python backend about the variables in scope and the code in the notebook.
The call returns a Python object that renders the HTML and Javascript of \sys's interface (the object has a \code{\_repr\_html\_} method for Jupyter).

\begin{figure}[b]
\centering
\begin{minipage}{0.78\columnwidth}
\vspace{-\intextsep}
\begin{notebooklisting}\begin{lstlisting}
def bar(
   self,
   x: float | ArrayLike,
   height: float | ArrayLike,
   width: float | ArrayLike = 0.8,
   bottom: float | ArrayLike = 0,
   *, # Subsequent args are keyword-only
   align: Literal["center", "edge"] = "center",
   color: ColorType | Iterable[ColorType] = ...,
   ...
   **kwargs: RectangleProps,
) -> BarContainer: ...	
\end{lstlisting}\end{notebooklisting}\vspace{-0.9em}
\end{minipage}
\caption{A subset of type annotations for \code{ax.bar}}
  \label{fig:types-ax-bar}
\Description{
A subset of type annotations for the "ax.bar" call.
}
\vspace{-0.35em} 
\end{figure}
\mypara{Properties Panel - GUI from Type Annotations}
To determine possible function arguments for the Properties panel and suggested values for each, \sys runs the MyPy~\cite{MyPy} type checker on the notebook code to find function arguments types. Consider the type annotations in \autoref{fig:types-ax-bar} for the \mpl \code{ax.bar} function,
%
%
these become the possible arguments shown in the Properties panel.
The types describe the possible values for each argument, as well as a possible default value (after the \code{=} sign). 
The default is used as the initial value when the user clicks to enable the argument,
and the type determines the suggestions in the dropdown.
For example, \code{bar}'s first argument \code{x} has type \code{float | ArrayLike}, meaning it may be either a \code{float} or an \code{ArrayLike} (\code{ArrayLike} means what NumPy considers array-like objects). For \code{float} and \code{ArrayLike}, \sys offers the suggestions \code{0.5} and \code{[1, 2, 3]} respectively. Any of the user's variables that are type-compatible with \code{float | ArrayLike} are also suggested, as well as depth one indexing into these variables, as shown in \autoref{fig:ex_making_bars} (middle).

\begin{figure}[h]
\centering
\includegraphics[width=0.85\columnwidth]{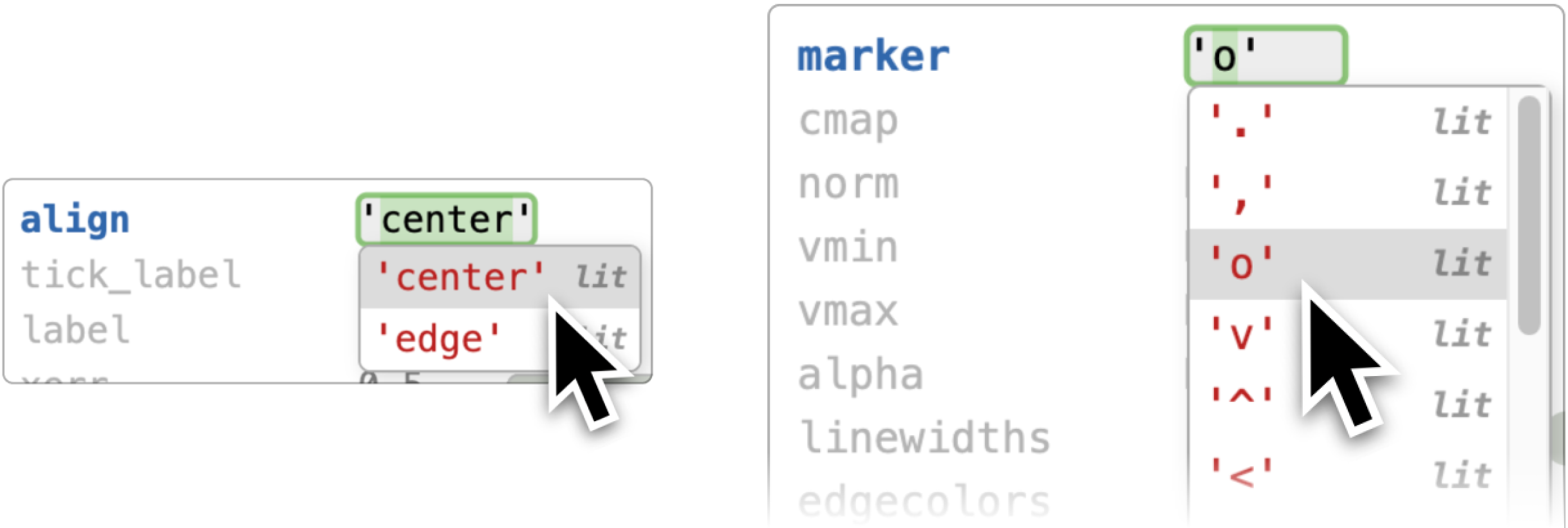}
\caption{The Properties panel is convenient for parameters with \code{Literal[...]} type annotations.
Options are live-previewed.
}
  \label{fig:suggestions_dropdowns}
\Description{
Left: property "align" highlighted with a dropdown showing 'center' and 'edge'.
Right: property "marker" highlighted with a dropdown showing many value suggestions for the property, cursor is over 'o'.
}
\end{figure}

As a second example, consider \code{ax.bar}'s \code{align} argument. Its type is \code{Literal['center', 'edge']}, that is, these two strings are the \emph{only} possible values,
so the suggestions dropdown shows exactly these two (\autoref{fig:suggestions_dropdowns}~left). Hovering the cursor over each previews the result on the plot: \code{'center'} centers the bars over the \code{x} values, whereas \code{'edge'} aligns the left edges of the bars on the \code{x} values. 
\code{Literal} annotations make working with such enum-like properties smooth in \sys: for example, to choose a shape for scatterplot points the user does not have to look up documentation in a separate browser tab, they can just click the \code{marker} argument and quickly preview the suggestions by scrubbing their mouse over each (\autoref{fig:suggestions_dropdowns}~right).



Certain other types are backed by UI widgets in the Properties panel. Floats and ints may be manipulated with a slider \raisebox{-0.2em}{\includegraphics[height=1em]{slider}}\Description{A horizontal slider}, booleans with a toggle switch \raisebox{-0.13em}{\includegraphics[height=1em]{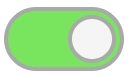}}\Description{A small on/off toggle switch}, and colors show a swatch \raisebox{-0.13em}{\includegraphics[height=1em]{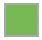}}\Description{A small green square}\ which opens an HTML5 color picker. Widgets are placed to the right of an ordinary text box so the user is also free to write arbitrary code.


\noindent

\sys bundles its own \mpl type annotations. We started from Microsoft's Python Type Stubs project~\cite{typestubs} and added and modified the types. For example, we might write a type \code{Literal['a',} \code{'b'] | str}, even though it is the same as just \code{str}, because \code{Literal['a',} \code{'b']} produces \code{'a'} and \code{'b'} as dropdown suggestions.

\begin{figure}[b]
\includegraphics[width=0.49\columnwidth]{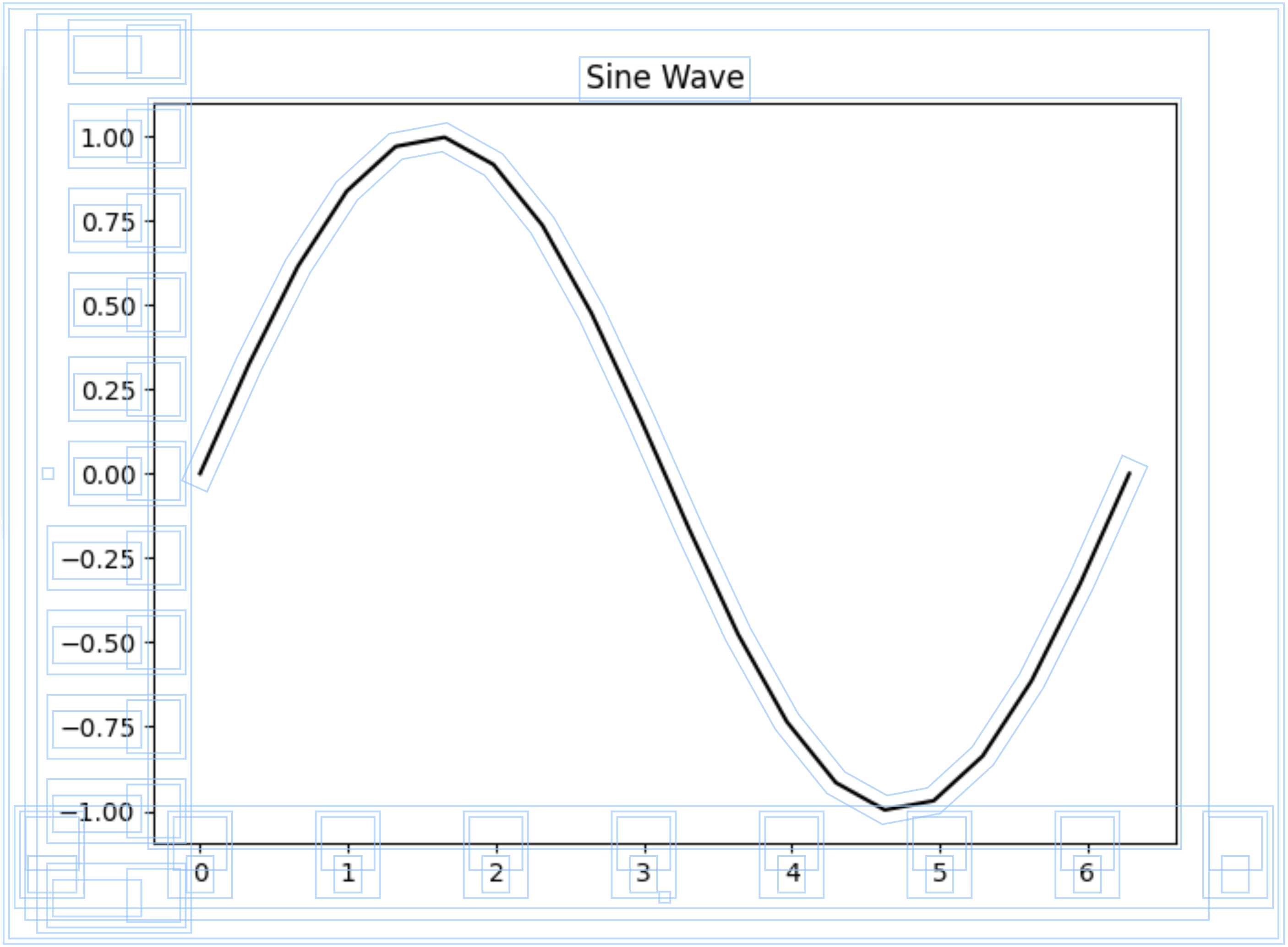}
\hfill
\includegraphics[width=0.49\columnwidth]{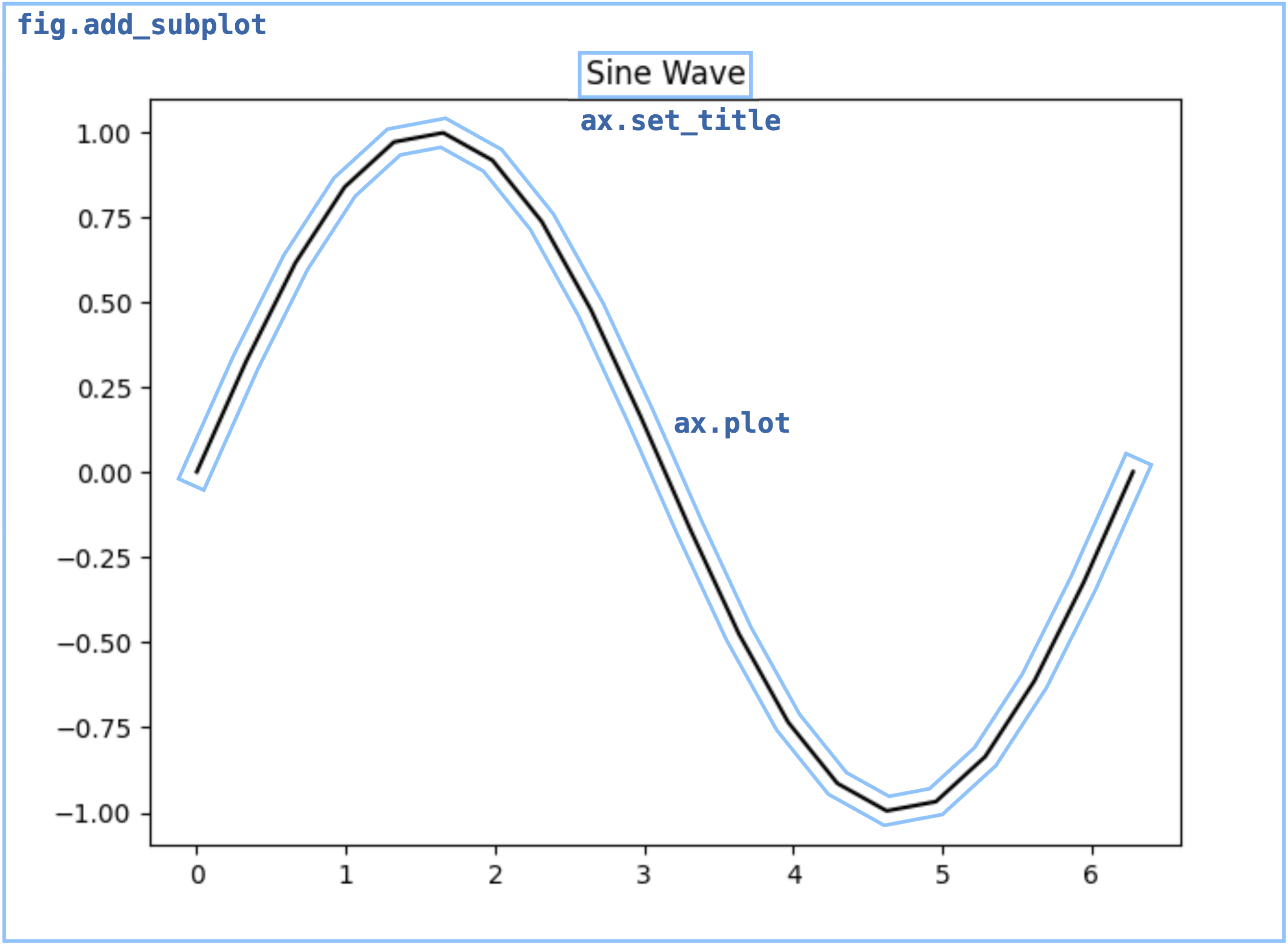}
\caption{Regions for all shapes in the scene graph (left, in blue) and those clickable to select a call in the user's code (right).}
\label{fig:scene-graph}
\end{figure}
\mypara{On-Plot Regions for User Calls}
Internally, \mpl figures are a scene graph of shapes. \sys needs to know where these shapes are, both to make appropriate shapes on the output clickable and to position the on-plot buttons (like \onplotbutton{ax.set\_xlabel}). \sys first walks the \mpl scene graph and converts shapes into polygonal outline regions (\autoref{fig:scene-graph}~left), leveraging the \code{shapely} Python library~\cite{shapely} to expand each region slightly for clickability.
Some shapes \emph{should} be clickable to select one of the user's function calls (\autoref{fig:scene-graph}~right), but \mpl's scene graph does not ordinarily remember which calls produced the shapes. To preserve this correspondence, \sys rewrites the user's code to tag values returned from functions with location metadata: \eg \code{text = ax.set\_title(...)} becomes \code{text = tag(ax.set\_title(...), 'ax.set\_title \#1')}, which sets the propery \code{text.\_came\_from} \code{= 'ax.set\_title \#1'}. Shapes from function calls now have a \code{\_came\_from} property which \sys uses so that clicking the shape will select the function call in the Layers and Properties panels.

\mypara{On-Plot Buttons}
Similarly, \sys positions on-plot buttons for adding new function calls, \eg \onplotbutton{ax.yaxis \btd} and \onplotbutton{ax.set\_title}. \sys has a hard-coded list of which functions should appear on which parts of the plot, \eg{} if the \code{ax} variable is an \code{Axes} object, \sys will make \code{ax.set\_ylim()} appear on the \code{ax.yaxis} shape. The challenge is figuring out that the relevant root object is named \code{ax}.
Naively, we could walk the scene graph to determine object names, but this will produce ugly code like \code{fig.axes[0].set\_title} instead of just \code{ax.set\_title}.
Instead, \sys{} walks from the \textit{local variables} to a depth of 4. 
If there are multiple paths to an object, \eg \code{axs[0]} and \code{fig.axes[0]}, \sys uses the shortest. The buttons on the plot, \eg \onplotbutton{axs[0].set\_title}, use these idiomatic names when adding new code.


\mypara{On-Plot Direct Manipulation}
\sys lets the user directly manipulate certain shapes on the plot to move or resize them.
Note that this is different from \eg{} iPyWidgets~\cite{iPyWidgets} where interactions only modify the Python state behind the scenes: \sys changes the code.
The parts of a shape that are draggable are determined by looking for specific argument names within the function's first ten arguments, such as ``\code{x}'' or ``\code{y}'' for the shape body, or ``\code{xmax}'', ``\code{right}'', or ``\code{width}'' for the right edge.
If the argument is currently a number, \eg \code{x=10}, then dragging the shape will change the number, otherwise dragging will add a number to the expression, \eg \code{x=var+10}.
This local code update scheme is similar to CapStudio~\cite{CapStudio}, there is no backpropagation deeper into the code as in other systems~\cites{sns-pldi,sns-oopsla,FusingDirectManipulationsIntoFunctionalPrograms}.
To determine how much a number should change per pixel of mouse movement, \sys tags shapes with the data limits of their containing axes and the axes size in pixels, to convert from pixels to the axes units.

\begin{figure}[t]
\centering
\begin{minipage}{0.4\textwidth}
  \begin{minipage}[c][0.7in]{1.2in}\raggedleft\includegraphics[height=0.6in]{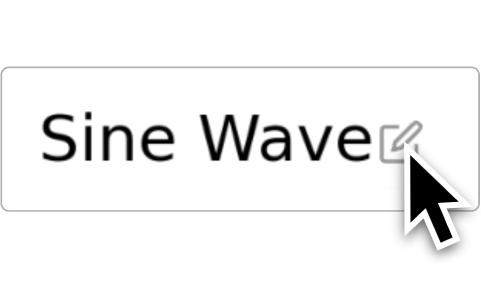}\end{minipage}
  \hfill
  \begin{minipage}[c]{0.2in}\includegraphics[width=\textwidth]{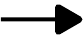}\end{minipage}
  \hfill
  \begin{minipage}[c][0.7in]{1.05in}\includegraphics[height=0.6in]{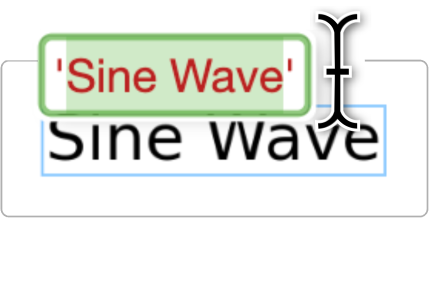}\end{minipage}
\vspace{-0.5em}
\end{minipage}

\caption{Changing a plot title directly on the plot.}
\label{fig:on-plot-dm}
\Description{
Plot title "Sine Wave" being changed directly on the plot.
Left: cursor is over the edit icon next to the title.
Right: text box is opened above the title showing the text "Sine Wave" for editing.
}
\end{figure}

The text for some items on the plot, \eg{} title text and axis labels, can be edited by clicking an adjacent edit button
without going to the Properties panel (\autoref{fig:on-plot-dm}).



\section{Evaluation}\label{sec:eval}

To understand the trade-offs of \sys's multi-modal approach, we conducted an IRB-approved two-part study---an exploratory lab study (``\studyone'') and a longer follow-up study (``\studytwo'')---focusing on when and why users switch between GUI, code, and AI.
Our methodology and findings are below.

We also used \sys to create charts in \autoref{fig:interaction_counts} and \autoref{fig:time_percentage} and plots from earlier drafts of the paper (\eg \autoref{fig:teaser}). Although not discussed here, the appendix at \plotteryDotOrg~\suppMaterialsRef includes those plots and our reflections on authoring them, as well as reflections on a plot too complicated for \sys, a narrated video of the walkthrough example, and the study replication package.

\subsection{User Study Methods}\label{subsec:user-study-methods}

\noindent Studies were in person and recorded with participant consent.

\mypara{Participants}
We recruited 10 participants for \studyone via institutional mailing lists and messaging platforms (5 women, 4 men, 1 undisclosed):
9 graduate students (6 in Computer Science, 1 in Data Science, 1 in Bioinformatics, and 1 in Electrical Engineering), and 1 Computer Engineering undergraduate.
Based on their self-reported comfort with \mpl, we classified 5 as experts (P2, P4, P5, P7, P9), and 5 as non-experts (P1, P3, P6, P8, P10).
%
Four participants further attended \studytwo (detailed below).
%
%

\begin{figure*}[t]
\begin{minipage}[t]{0.78\textwidth}
    \centering
    \includegraphics[width=\textwidth]{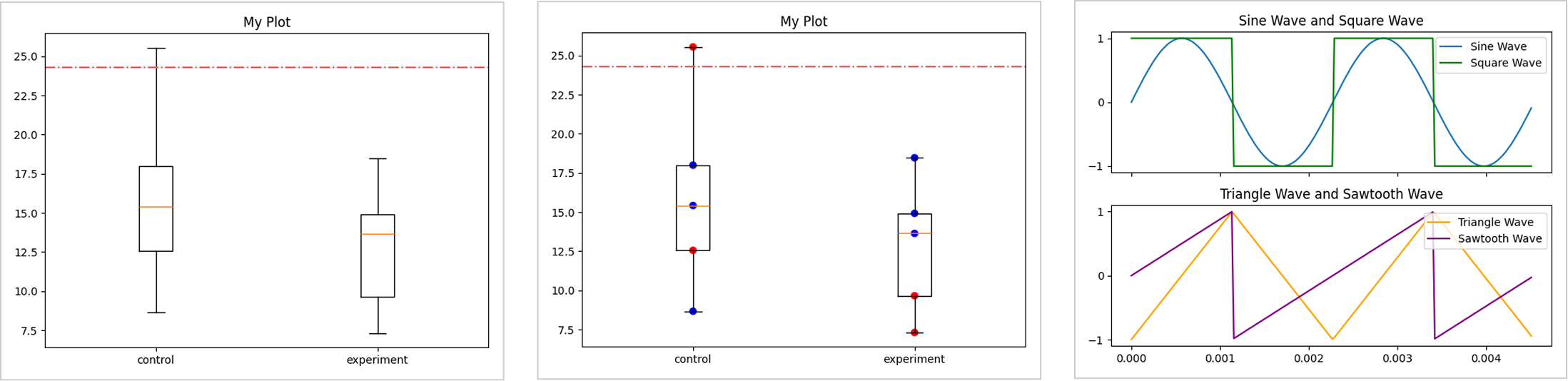}
    \caption{\studyone tasks. Task 2 extends the first, with GUI/AI alternately ablated. Task 3 is free use.}
    \label{fig:part_1_study_tasks}
\end{minipage}
\hfill
\begin{minipage}[t]{0.19\textwidth}
    \centering
    \includegraphics[width=\textwidth]{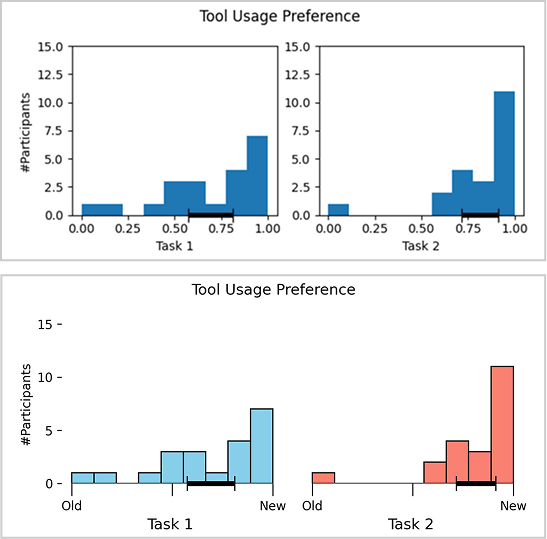}
    \caption{\studytwo primer.}
    \label{fig:part_2_refresher_tasks}
\end{minipage}
\end{figure*}

\mypara{\studyone (60min)}
Participants did a 10 minute tutorial on all \sys modalities using a bar chart like that in \autoref{sec:overview-example}, then completed three timed tasks (\autoref{fig:part_1_study_tasks}),
each in one of three conditions (internet search always allowed):
\begin{itemize}[leftmargin=2.6cm]
    \item[(C1 \AI)] No GUI use allowed, 10min; 
    \item[(C2 \DM)] No AI prompting allowed, 10min;
    \item[(C3 \Mixed)] All \sys modalities allowed, 15min.
\end{itemize}
The condition order
was randomly C1-C2-C3 or C2-C1-C3.
%
%
Finally, participants completed a survey and a semi-structured interview to reflect.

\begin{figure*}[t]
\includegraphics[width=\textwidth]{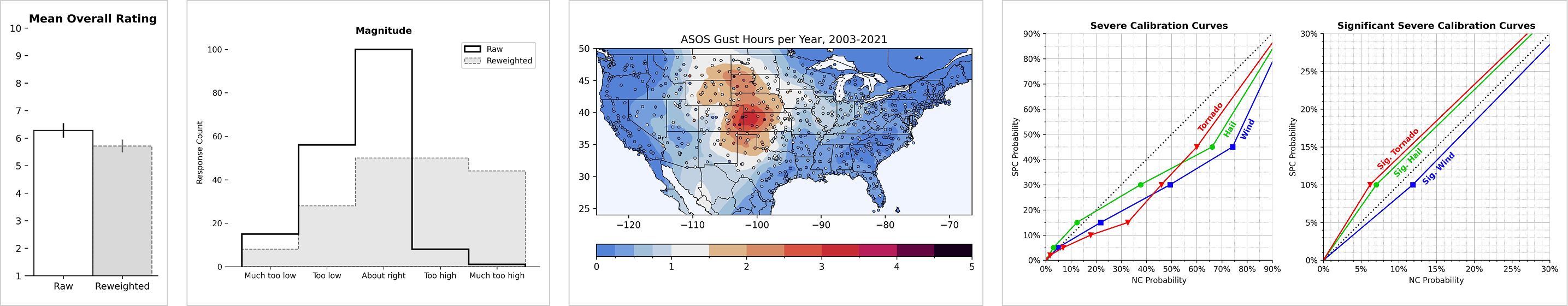}
    \caption{
    The possible C3 \Mixed tasks for \studytwo, adapted from real-world plots.
    The tasks cover the topics of the 20 most-visited non-tutorial pages on matplotlib.org.
    The map task used \code{imshow} and a scatterplot rather than geospatial libraries.
    }
	\label{fig:part_2_study_tasks}
\end{figure*}

\mypara{\studytwo (120min)}
Four participants (P4, P7, P9, and P10, denoted for \studytwo as \P{4}, \P{7}, \P{9}, and \P{10}) attended \studytwo for more extended use of \sys to recreate
real-world 
plots adapted from the first author's work. The plots collectively cover the topics of the 20 most-visited non-tutorial \mpl documentation pages~\cite{matplotlibPageStats}.
Participants completed a 25-40 minute, experimenter-facilitated \sys refresher on one plot (\autoref{fig:part_2_refresher_tasks}), authoring half without the GUI and the other half without AI (C1 and C2, order counterbalanced),
then independently attempted the remaining four plots (\autoref{fig:part_2_study_tasks}, order counterbalanced)
using all modalities freely (C3 \Mixed) limited to 40 minutes on each task. We did not expect participants to complete all four tasks given their complexity, and indeed only P7 completed a task on time.
For the final 10 minutes, they participated in a semi-structured interview.
\mypara{\diff{}{Analysis}}
For \studyone, we performed \emph{quantitative} analysis as follows: 
(1) task success---whether a participant completed the plot within the time limit---for all tasks;
(2) usage counts and durations for each modality
for the C3 \Mixed task;
(3) modality switch counts and their cause (either failure from the previous interaction or just user preference) for the C3 \Mixed task.
We further did a \emph{qualitative} thematic analysis of \sys-relevant behaviors and comments regarding modalities, benefits, breakdowns, and user desired features.

For \studytwo,
we \emph{quantitatively} measured (1) modality usage counts and durations for the C3 \Mixed tasks, and (2) which code originated from which modality, with \emph{qualitative} analysis of the code's purpose (\eg add/duplicate/tweak). We also scrutinized the videos to characterize interaction breakdowns.

\subsection{User Study Results}\label{subsec:user-study-results}

\autoref{tab:takeaways} summarizes the takeaways from our evaluation: tasks that might suit each modality, the downsides of each, and findings on modality use. We discuss these below.

\begin{table}[t]
	\footnotesize
	\centering
	\caption{Evaluation takeaways.}\label{tab:takeaways}		
  \setlength{\tabcolsep}{3pt}
  \newenvironment{cellList}
  {\vspace{-1em}\begin{itemize}[leftmargin=*,label=$\cdot$,labelsep=2pt]}
  {\end{itemize}}
  \begin{tabular}{c >{\raggedright\arraybackslash}p{0.312\columnwidth} >{\raggedright\arraybackslash}p{0.528\columnwidth}}
 \toprule
      & \textbf{Suited For}                                         & \textbf{Downsides} \\
\midrule
\textbf{\color{GUIColor}{GUI}}   &\begin{cellList}\item Straightforward visual changes.\end{cellList}                           &\begin{cellList}\item Overwhelming number of options.\item Less helpful than AI.\item Still requires API knowledge.\end{cellList}  \\ \midrule
\textbf{\color{CodeColor}{Code}}  &\begin{cellList}\item Quick and small edits.\item Duplicating code.\end{cellList} &\begin{cellList}\item Requires knowing the plot code and \mpl API.\end{cellList}      \\ \midrule
\textbf{\color{AIColor}{AI}}    &\begin{cellList}\item Starting tasks.\item Generating code when unsure of API.\end{cellList} &\begin{cellList}\item Can require validation.\item Sometimes need \mpl vocabulary.\end{cellList}\\
\midrule
\midrule
\multicolumn{3}{l}{
\begin{minipage}{0.97\columnwidth}
\textbf{Modality Use}:
\vspace{0.8em}
\begin{cellList}
  \item Participants exhibited varying strategies for modality use \emph{frequencies}, but usually spent the \emph{most time} with AI and produced \emph{most code} through AI.
  \item Participants performed $\sim$3 interactions in a modality before switching.
  \item Besides scrolling, it was easy to switch modalities.
  \item Live feedback is a key part of fluidity.
\end{cellList}
\end{minipage}
} \\
\bottomrule
\end{tabular}
\Description{Evaluation takeaways.}
\end{table}

\newcommand{\labeledtriplot}[2]{%
    \begin{tikzpicture}
        \node[inner sep=0pt] (img) {\includegraphics[height=2.8cm]{#1}};
        \node[
            anchor=north west,
            fill=black!5,
            rounded corners=2pt,
            inner sep=3pt,
            font=\scriptsize\bfseries
        ] at ([xshift=-3pt,yshift=3pt]img.north west) {\textnormal{\textcolor{black!70}{#2}}};
    \end{tikzpicture}%
}

\begin{figure*}[b]
\begin{minipage}[t]{\columnwidth}
    \centering
    \labeledtriplot{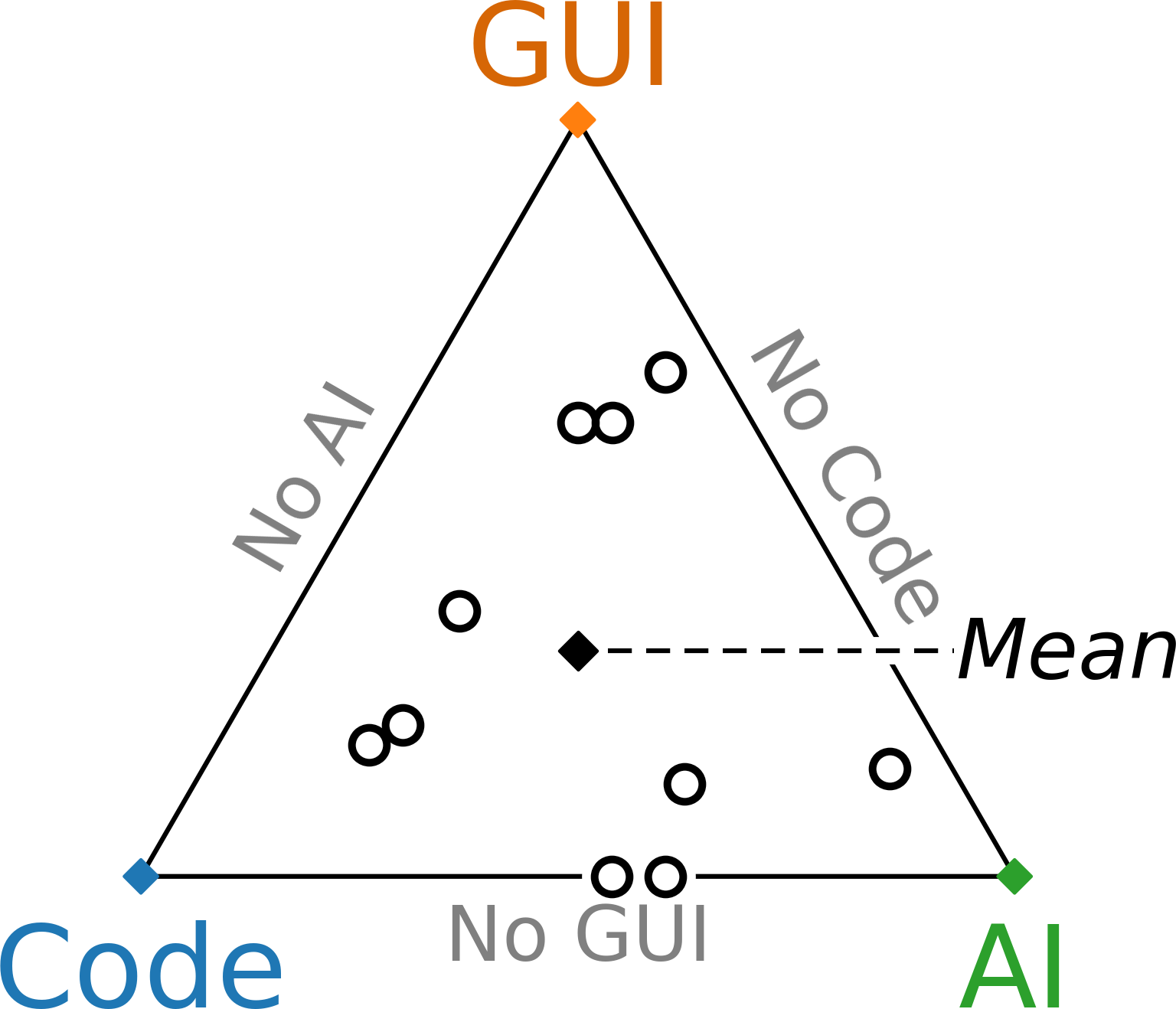}{\sffamily\studyone}
    \hspace{2.2em}
    \labeledtriplot{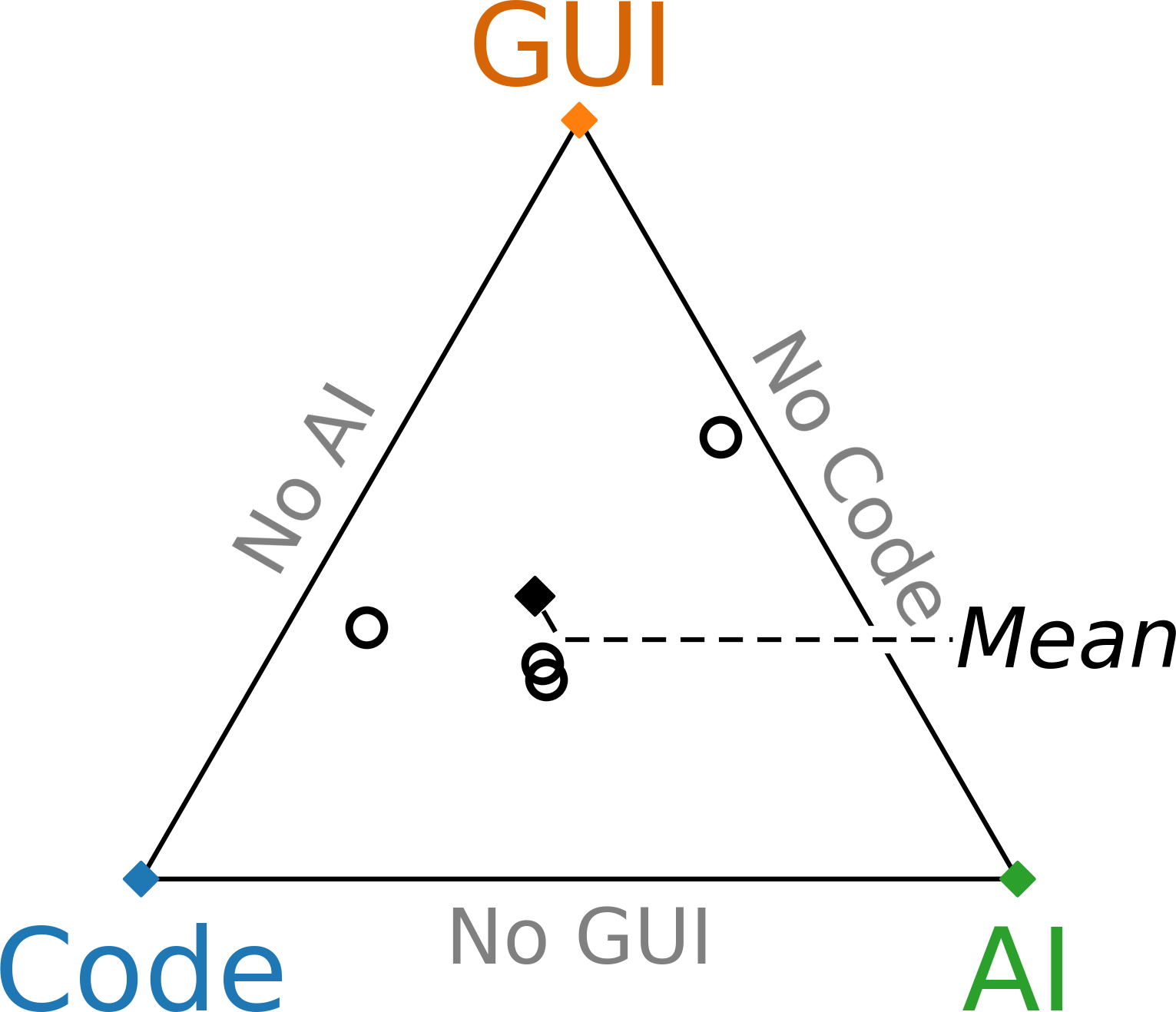}{\sffamily\studytwo}
    \\[0.3em]
    \caption{Relative \textit{number of interactions} in each modality. 
	Individual patterns vary but overall usage is balanced.}
    \label{fig:interaction_counts}
\end{minipage}
\hfill
\begin{minipage}[t]{\columnwidth}
    \centering
    \labeledtriplot{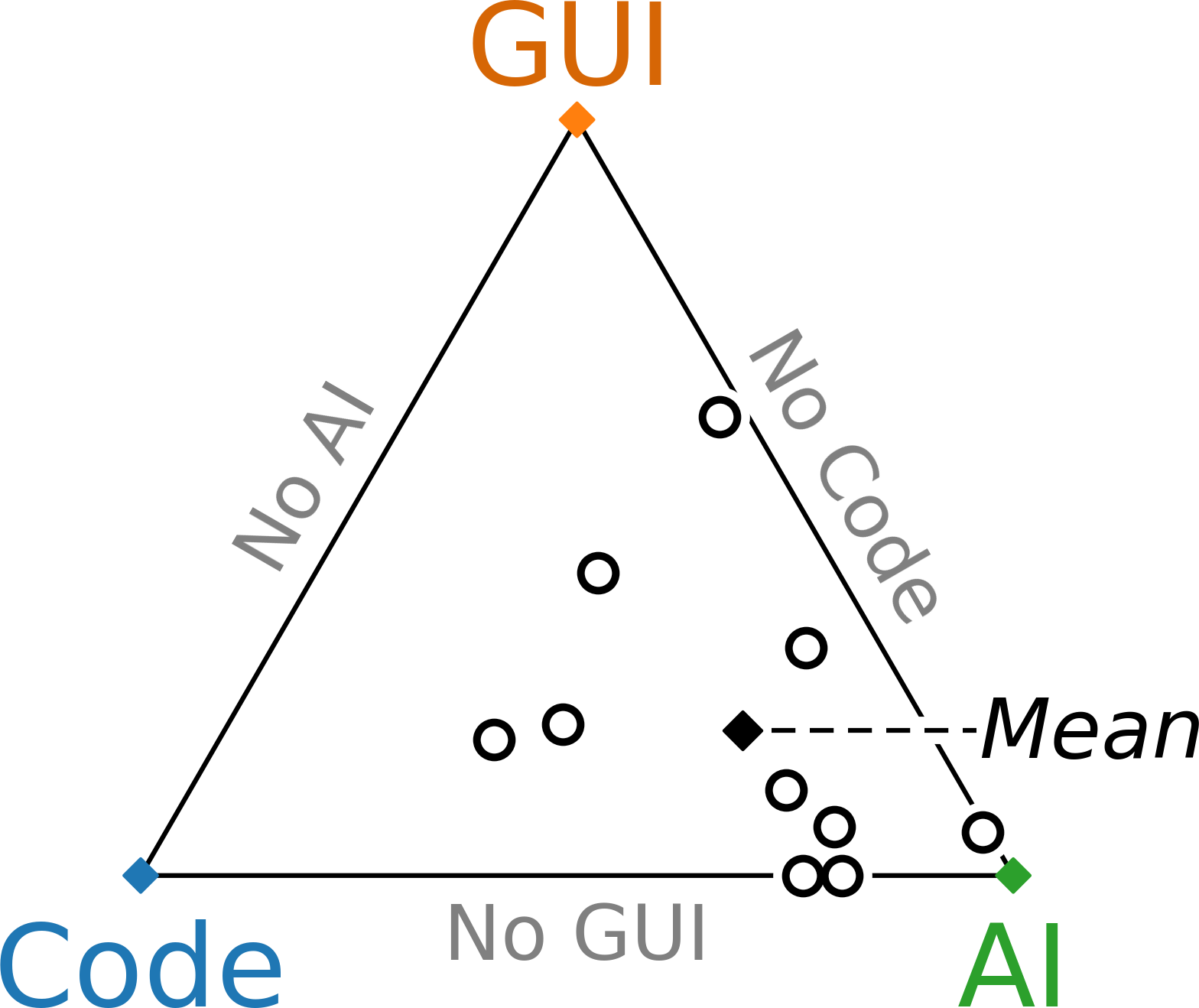}{\sffamily\studyone}
    \hspace{2.2em}
    \labeledtriplot{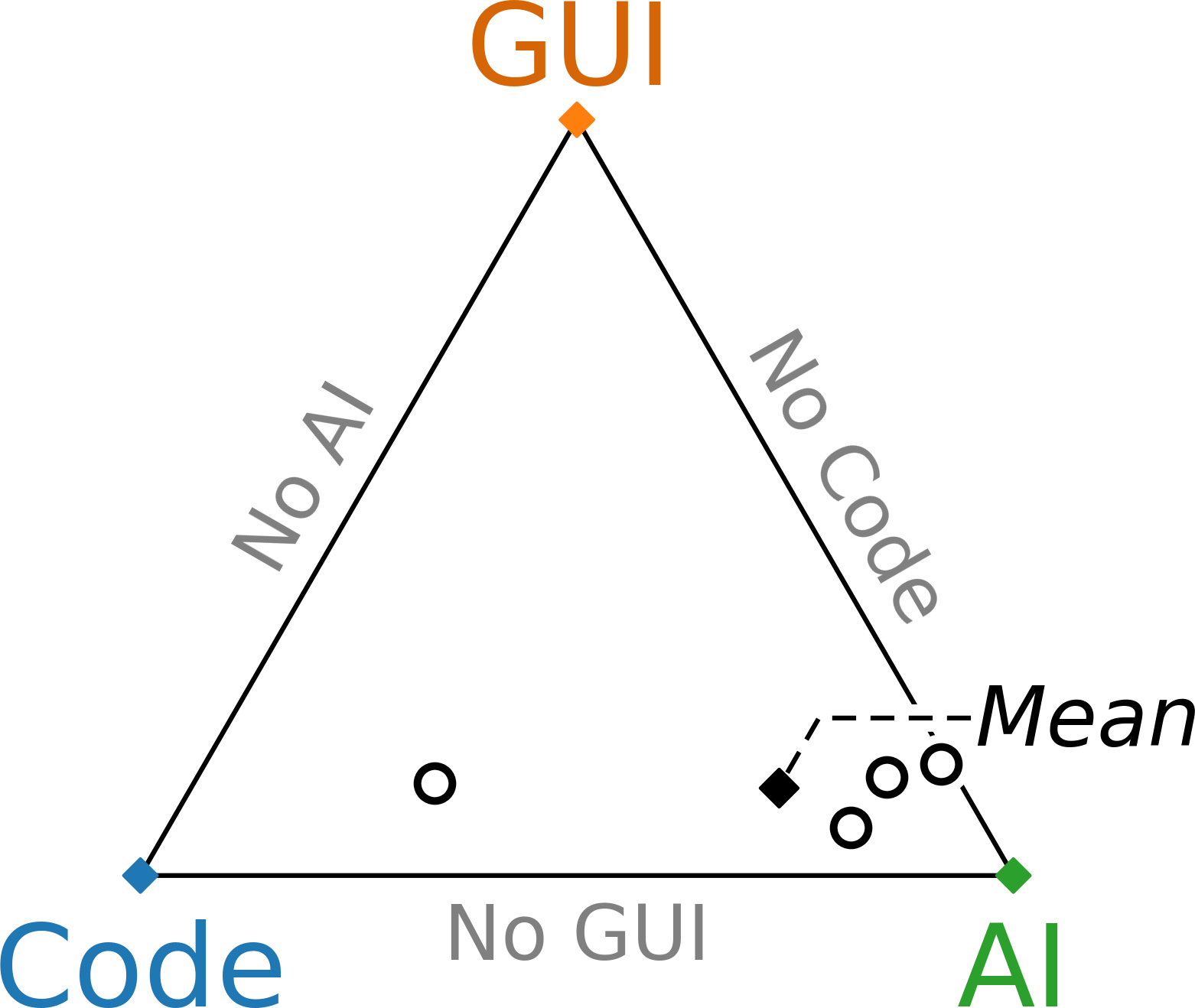}{\sffamily\studytwo}
    \\[0.3em]
    \caption{Proportion of \emph{interaction time} in each modality. For both parts, AI occupies a majority of time and GUI the least.}
    \label{fig:time_percentage}
\end{minipage}
\end{figure*}

\mypara{Modality Switching Was Fluid}
By usage \textit{counts} (\autoref{fig:interaction_counts}), the relative usage of the three modalities was balanced overall, though preferences varied. In \studyone, three participants favored the GUI (P1, P6, P9), three favored code (P3, P5, P10), and four favored AI (P2, P4, P7, P8); two participants never touched the GUI (P2, P7) and two almost never touched the code (P4, P6). 
\studytwo showed similar patterns of balanced overall usage and individual preferences.
Within that balanced usage, mode switching was fluid.
In \studyone, participants performed an average of 3.1 interactions in a single modality before switching, and 2.6 in \studytwo.
For comparison, every interaction in a random modality would average to 1.5 interactions between switches,
so $\sim$3 interactions at a time indicates
some bias to continue in the same modality, but switching modalities was nevertheless easy, a finding also supported qualitatively.
Participants found the multi-modal integration \emph{``seamless''} (P1) and \emph{``dynamic''} (P5): \emph{``I didn't spend too much mental effort figuring out [which to use]''} (P1), and seven participants (P2, P3, P4, P5, P8, P9, P10) highlighted the AI's in-situ responses, compared to using tools like ChatGPT in a separate tab.
Because of the size of \sys's GUI, we did notice a lot of scrolling, particularly by \P{7}, \P{9}, and \P{10}, but no participant mentioned it interfered with fluidly mode switching, perhaps because scrolling is normal in Jupyter~\cite{groundedJupyter}.
%

Feelings of fluid interaction may depend on \emph{live feedback}.
In \sys, GUI changes are immediately previewed and the code cell is automatically re-run upon AI-generated code.
We asked how participants perceived \sys's multi-modal interaction, and several (P3, P4, P5, P9) highlighted this liveness.
%
%
Participants appreciated
\emph{``immediately seeing the effect [of AI-generated code]''} without copying or pasting (P3),  
\emph{``[seeing] these things happen in real time [alongside] the underlying code''} (P9),
and \emph{``[feeling] easier to explore different options''} (P5) compared to the lack of liveness in code editing, in which one would \emph{``have to rerun the code''} (P4, P5).
%
As additional evidence,
the first author accidentally ablated the live feedback halfway through creating one of the figures in the Appendix (from a bug not encountered in the studies), which forced them to manually re-run the cell to see the results of every GUI or AI edit.
With the clunkier interface, the feeling of fluidity was lost.
%
%
These results suggest that liveness is a necessary piece for fluid interaction in \sys.


\mypara{The \textcolor{GUIColor}{GUI} may be suited for straightforward visual changes, but can overwhelm and helps less than AI}
Based on their comments (\studyone) and usage (\studytwo),
participants found direct manipulation in the GUI useful for many kinds of edits,
such as \emph{``repositioning the legend [and] resizing the bars''} (P1, \P{4}, P6), \emph{``[changing] plot titles [and] tick labels''} (P2, \P{4}, \P{7}, \P{9}), or \emph{``moving things around''} in general (P3, P4).
The moment P6 learned about on-plot direct manipulation in the tutorial when using it to resize bars, they exclaimed: \emph{``Oh, my ...! The amount of times I have tried to figure out how to make the bars skinnier...''}
%
%
Participants also found the GUI useful for adding/removing/restyling a visual component (P4, P6, \P{7}, \P{9}, \P{10}), data-relevant visual changes (P1, P6), or exploring property changes---especially colors (P1, P4)---when they were \emph{``not sure what the exact [argument or variable] would be''} (P5) and did not want to search the documentation (P2), or simply wanted to learn about \emph{``how each thing works in \mpl.''}
Despite the GUI's support for these \emph{straightforward} visual changes, participants struggled to complete their tasks with it: only 3/10 \studyone participants successfully completed their task in the \DM condition with AI ablated (P2, P5, P9), while all participants completed their \AI task without the GUI.
These \DM failures surfaced a common pattern: the sidebars show some of \emph{what} could be changed about the visual components, but not \emph{how} to achieve the exact desired changes.
Indeed, both \mpl experts (P2, P4, P7) and non-experts (P3, P6, P8, P10) felt the need to search online to figure out how to use \sys to solve the task in \DM, as the GUI (compared to AI) sometimes could not help translate user intents to \sys actions (P2, P4, P6).
Additionally, some found the GUI widgets overwhelming (P2, P3, P6) with \emph{``too many things at once''}, most of which they did not know (P3), and inefficient for larger changes when compared to AI (P2, P3, P5, P6).
We classified whether mode switches in \studyone were from goal failure or just user preference. We found no one used the GUI as their fallback when failing in code or AI, and 13 of 27 switches from the GUI are failure-driven (48\%), in contrast to 4 of 32 switches away from AI (13\%) and 7 of 25 switches away from code (28\%).
This does not mean the GUI is unhelpful—52\% of GUI interaction sequences ended in success—nor that the AI can subsume the usefulness of the GUI—the GUI can \eg help you discover unfamiliar plotting functions and arguments (\P{9})—but it does reveal that the GUI often presents an overwhelming number of options which users may not know how to navigate.

\mypara{Despite requiring \mpl fluency, direct \textcolor{CodeColor}{code} editing may be suitable for quick/small edits and duplication}
All participants in both parts edited some code.
Based on comments (\studyone) and usage (\studytwo), participants found code editing suitable for quick and small changes (P1, P2, P4, \P{4}, P6, P7, \P{7}, P8, \P{9}, P10, \P{10}), like tweaking parameters previously introduced by AI or the GUI (\P{4}, \P{7}, \P{9}, \P{10}).
Code editing may also be suitable for repeating existing code (P1, P2, P3, P5), which might have been previously generated by AI (\P{9}, \P{10}). Some in \studyone also edited code to gain control over \sys's AI when in doubt (P3, P4, P9).
Still, participants felt code editing requires familiarity with the code (P1, P2, P5, P9) and is \emph{``most straightforward if you have already memorized the correct keywords''} (P5).

\mypara{\textcolor{AIColor}{AI} may be broadly helpful for starting tasks and generating code when unfamiliar with the API, but nevertheless can still require knowing \mpl vocabulary}
\autoref{fig:time_percentage} shows that, by time, AI was the most common modality for 6/10 participants in \studyone and 3/4 in \studytwo.
AI also produced more code than other modalities for 8/10 participants in \studyone and 3/4 participants in \studytwo.
Participants reached for the AI to make initial progress on a task: 3/4 \studytwo participants always started their tasks with AI, and although 6/10 \studyone participants started in the GUI, 5/6 of those interactions failed, and it was the AI that led to initial task progress for 8/10 participants.
Indeed, participants commented that AI helped initiate a task (P1, P4, P5, P6, P8), generate unfamiliar code (P1, P4, P5, P6, P7, P8, P10), accelerate typing many lines (P2, P7, P10), and replace internet searches (P3, P6).
%
%
But, successful prompting may sometimes depend on knowing \mpl vocabulary.
P8, a non-expert, timed out while failing to restyle X-axis ticks with the AI (after already failing with the GUI): some prompts had typos, which neither the AI nor they recognized, and some prompts led to unexpected changes that set back their progress.
This single observation in \studyone was exacerbated in \studytwo, where the colorbar on the map task explicitly called for two \mpl concepts, but \P{4} and \P{7} still failed to progress as they repeatedly prompted the AI with descriptions of the plot's appearance rather than \mpl-specific terminology.

\section{Discussion}\label{sec:discussion}

\sys's goal is to seamlessly augment plot code editing with GUI (and AI) interaction
to showcase an ideal plotting workflow that is both as ``GUI-like'' and ``code-like'' as possible.
Specifically, our design goals (DGs) are to provide:
\begin{enumerate}[leftmargin=*]
\item \textbf{Familiarity} by imitating a traditional graphics editor;
\item \textbf{Fluid iteration} via direct manipulation;
\item \textbf{Predictability} in how GUI actions become code;
\item \textbf{Compatibility} with code’s benefits; \textit{and}
\item \textbf{Fluid mode switching} between GUI, code, and AI.
\end{enumerate}
\noindent
Below we discuss the degree to which we feel we succeeded, what generalizes beyond plotting, and possible future work.

\mypara{(DG1) A familiar GUI in the AI era}
While users seemed to understand ``Calls = Layers'' and ``Arguments = Properties'' without trouble,
study participants still spent relatively little time with the GUI (\autoref{fig:time_percentage}) and most code for most participants came from AI.
AI reliance may have been exacerbated by our study design:
the tasks were well-defined reproduction tasks, whereas real-world authoring may involve more exploring of visual style and more exploring of the GUI interface.
Even so, most participants did use the GUI (\autoref{fig:interaction_counts}), and our study does not suggest that AI prompting is a full replacement for graphical interaction. Participants used and appreciated the GUI for straightforward visual changes---it is simply nicer to drag a font size slider until it looks right than to prompt ``title bigger'' and ``title a little smaller'' over and over.
Indeed, major AI coding tools are introducing direct manipulation for tweaking code-generated visual designs~\cites{CursorBrowserVisualEditor,v0DesignMode},
albeit applied to code by AI and without any interface for binding properties to existing code (\autoref{fig:ex_making_bars}).
Nevertheless, the utility of AI in our user study highlights that graphical interfaces,
despite promising discoverability~\cite{DirectManipulation},
do still require learning in a way that AI prompting does not.
\sys's long function list in the \onplotbutton{ax \btd} menu and long  property lists were problems.
After \studyone, we added a search box to the \onplotbutton{ax \btd} menu, but could also add search to the Properties panel or could organize related properties into tabs and groups.

A recent alternative to an overwhelming GUI is dynamic widgets~\cite{DynaVis}, mini-UIs generated in response to AI prompts like in DynaVis~\cite{DynaVis} or BISCUIT~\cite{BISCUIT}. Although dynamic widgets require prompting first---there is no GUI before an AI interaction---we speculate that such transient widgets might be better than a static GUI in two scenarios: (1) changing uncommon properties, or (2) changes that require multiple GUI steps or are not possible in the GUI.
A possible hybrid between dynamic and static GUIs might be to treat the AI prompt box as a search box as well: as the user types it might focus the relevant part of the GUI (similar to CodeMend~\cite{CodeMend}) or after dispatching an AI prompt it might also open the relevant part of the GUI for further tweaking.

\mypara{(DG2) Fluid iteration}
To give pleasant feelings of direct manipulation~\cite{DirectManipulationCogSci},
\sys aims to embody the ``temporal, spatial, [and] semantic'' immediacy suggested by Ungar et al. \citet{ZStepTimeTravelDebugging}.
\emph{Temporally,} GUI or AI changes are visualized instantly—even just scrubbing through a suggestions list instantly previews each item.
\emph{Spatially,} GUI interactions are close to plot elements—to move the legend, the user drags it; to edit the title, the user clicks it.
\emph{Semantically,} \sys imitates a traditional editor—reducing semantic distance to a familiar interface—while still producing idiomatic code.
Further immediacy improvements are possible.
\sys lags on larger plots and multi-plot scenarios. Automatically reducing render resolution might help. Also, users often undo AI results and re-prompt: with a faster AI, \sys could instead preview the prompt result \emph{before} the user hits enter.
%

\mypara{(DG3\&4) Predictable, compatible code generation}
To make GUI actions predictable, we aimed for a straightforward linkage between code and GUI by
treating the user's function calls as a ``Layers panel'' and their arguments as a ``Properties editor''. Although simple in retrospect, this choice was not obvious.
The first prototype of \sys followed a different design, like Pylustrator~\cite{Pylustrator},
that organized shape selection and property editing around \mpl's internal scene graph rather than the user's code.
For example, setting the \mbox{X-axis} label and bolding it would involve editing the internal \code{ax.xaxis.label} shape, producing the code:

\begin{center}\begin{minipage}{0.8\columnwidth}\begin{notebooklisting}\begin{lstlisting}
ax.xaxis.label._text = 'My Label'
ax.xaxis.label._fontproperties._weight = 'bold'
\end{lstlisting}\end{notebooklisting}\end{minipage}\end{center}

\noindent
But with the ``Calls = Layers'' and ``Arguments = Properties'' commitment, \sys instead produces the idiomatic code:

\begin{center}\begin{minipage}{0.8\columnwidth}\begin{notebooklisting}\begin{lstlisting}
ax.set_xlabel('My Label', fontweight='bold')
\end{lstlisting}\end{notebooklisting}\end{minipage}\end{center}

\noindent
Moreover, users can \emph{write} that idiomatic code and have GUI support for editing it.
\sys's compatibility with user-written code extends further, to a key reason users plot with code: producing multiple plots. For example, the plots in \autoref{fig:interaction_counts} and \autoref{fig:time_percentage} are one function that is called four times. We can use \sys to restyle all four plots simultaneously.

\mypara{(DG5) Fluid mode switching, but not a silver bullet}
Adding more interaction modalities into a system, to gain the ``best of all worlds'', also loads a new burden on the user: they must choose which modality to use for each action~\cite{WhyProgrammingByDemonstrationSystemsFail}.
Switching costs can add friction, a pitfall we hoped to avoid.
Our attempt to provide \emph{``seamless''} (P1) and fluid~\cite{mage} interaction seems successful based on the study results. Besides persistent scrolling (which users seemed okay with) we didn't notice troubles switching modalities.
Most participants used all the modalities,
switching freely after $\sim$3 interactions on average.
%
But, seamless multi-modality is not a panacea for creating \mpl plots.
While each modality has advantages (\autoref{tab:takeaways}), these advantages do not always compensate for the others' failings.
With AI misinterpreting some prompts,
we observed instances where users floundered despite the presence of all three modalities.
Although improvements to the GUI and the AI models (\eg a vision LLM) might help,
multi-modality is not (yet) a silver bullet.
The \mpl API is complicated and reading documentation is still sometimes necessary.

\mypara{Better Data Binding GUI}
In \mpl, some kinds of data binding must be encoded manually by the user.
For example, in \studyone task 2 (\autoref{fig:part_1_study_tasks} middle) participants needed to transform a list of successes/failures into a list of blue/red colors.
Constructing such property lists is common in \mpl,
so GUI support for it, perhaps by simple embedded spreadsheets as in Lyra~\cite{Lyra}, is the most important missing feature we would like to add to \sys.

\mypara{Plotting libraries beyond \mpl}
\sys's ``Calls = Layers'' and ``Arguments = Properties'' GUI design is well-suited for imperative visualization libraries like \mpl that involve a sequence of function calls, each with a potentially long list of arguments.
The popular \mpl wrapper Seaborn~\cite{Seaborn} shares similar code structure, so \sys's design could be adapted to it given appropriate type annotations and updated function lists for the on-plot buttons.
D3~\cite{D3} and Altair~\cite{Altair} instead involve \emph{multiple chained} function calls to set object properties. To support this, we could treat a whole chain of function calls as a single selectable layer \emph{group}, with the Properties panel showing a vertical stack of panes, one for each call in the chain, for editing each call's properties.

\sys targets \emph{imperative} plotting. For \emph{declarative}, grammar of graphics (GoG)-based plotting~\cite{GrammarOfGraphics} the systems mage~\cite{mage}, Ivy~\cite{Ivy}, and Blace~\cite{Blace} all offer code editing alongside a GUI.
But, these GoG GUIs operate on the visualization grammar JSON, rather than on ordinary programmatic code, which limits their compatibility with common code features like loops, functions, or even just using a variable in scope. A potential next step for these GUIs is to consider how they can more thoroughly integrate into code-based analysis environments like Jupyter notebooks.

\mypara{Code-GUI inspectors beyond visualization}
Organizing the GUI around code structure opens the door to applying \sys's design to other code-based domains beyond visualization.
In particular, the typed-based Properties panel is not specific to \mpl.
The panel is similar to how IDEs show a type-based autocomplete as the user begins typing an API call.
But, the Properties panel instead makes the autocomplete a \emph{persistent} widget.
Ordinary IDEs for regular code could show a persistent panel of the arguments for the function call at the mouse cursor.
Prong~\cite{Prong} has a version of this that works for JSON-based languages, but type-driven GUIs can generalize beyond JSON or \mpl to ordinary programming.

\section{Conclusion}\label{sec:conclusion}

We designed and implemented \sys, a Jupyter notebook extension that integrates code editing with GUI support for creating plots with \mpl.
\sys imitates a traditional graphical editor by offering direct manipulation of shapes on the plot, and treating \mpl function calls as ``layers'' in a Layers panel and call arguments as ``properties'' in a Properties panel.
Edits are live, immediately reified in the code.
A prompt box lets users ask AI to modify the plot in-place.
A small, two-part exploratory user study indicates suitable usages for each modality—GUI for straightforward visual changes, code for quick edits and duplication, and AI for getting started and working around API ignorance—and showed that \sys enabled seamless, interchangeable use of these modalities.
%
%
We hope \sys's design sets a benchmark for future plot authoring and multi-modal programming tools that augment code's benefits—code's in-situ setting with data, flexible expressivity, and automation capabilities—with the addition of fluid GUI and AI interaction.

%

%

 \balance 

\bibliographystyle{IEEEtran}
\bibliography{references}

\clearpage
\nobalance
\section*{Appendix for ``\titletext''}

\begin{figure*}[b]
 	\centering
     \includegraphics[width=0.7\linewidth]{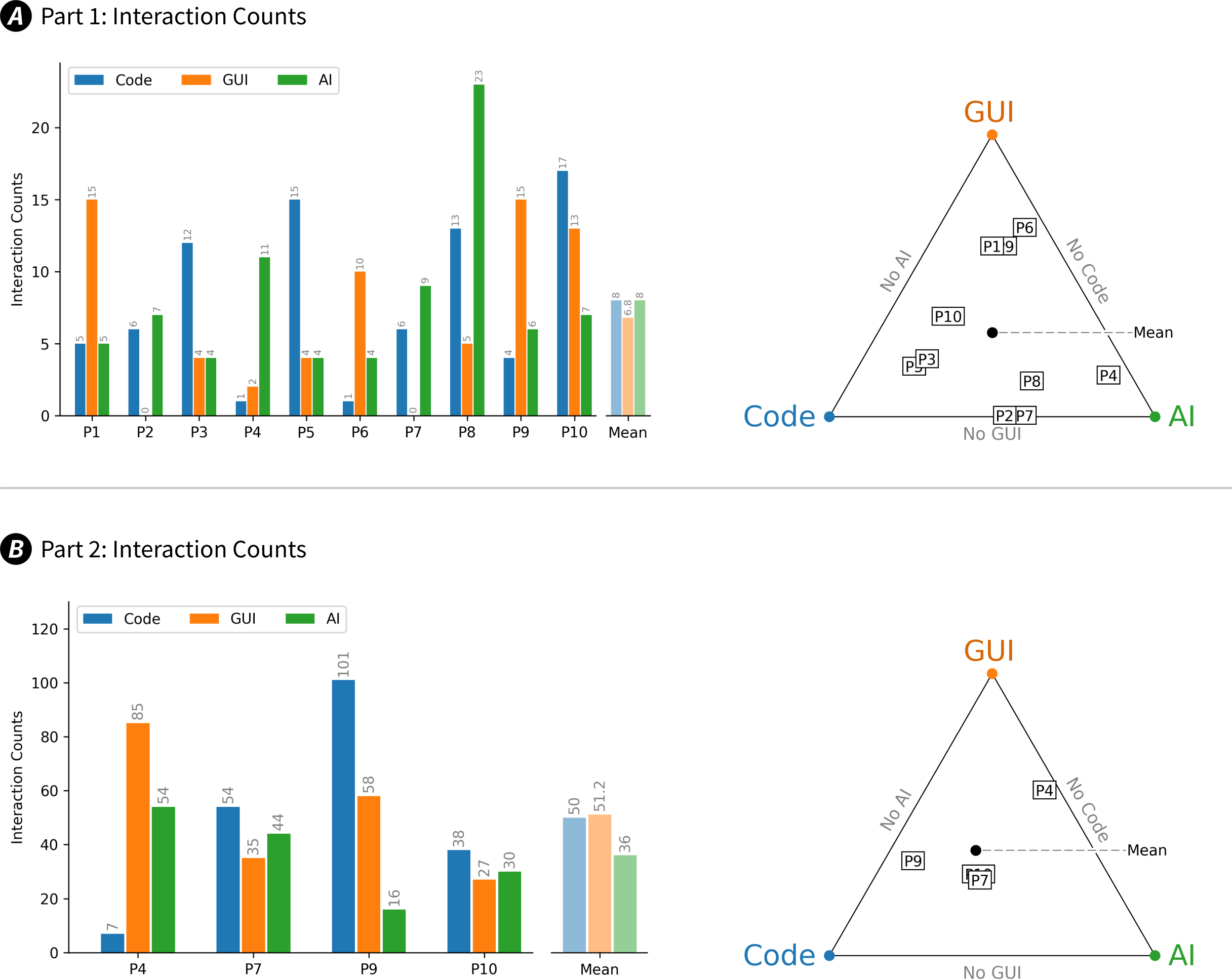}

 	\caption{
 	\emph{Interaction counts} when participants could use code, GUI, and AI freely: \circled{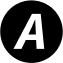} \studyone task 3 and \circled{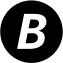} \studytwo tasks.}
 	\label{fig:old_interaction_counts}
 \end{figure*}
 
  \begin{figure*}[t]
 	\centering
     \includegraphics[width=0.7\linewidth]{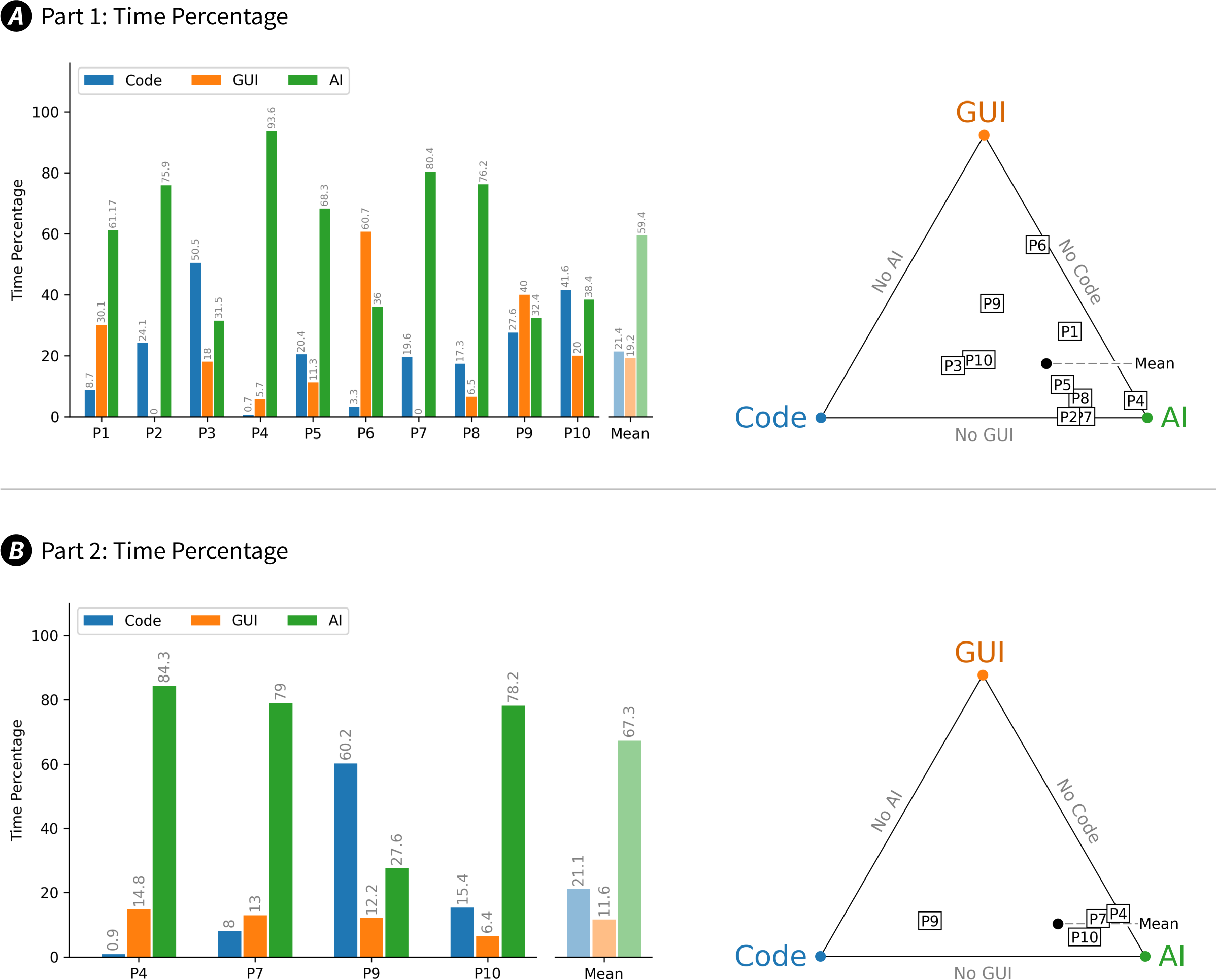}
 	\caption{
 	As in \autoref{fig:old_interaction_counts}, but percentage of \emph{interaction time} in each modality for \circled{A} \studyone task 3 and \circled{B} \studytwo tasks. For both parts, AI occupies a majority of interaction time and GUI the least.
 	}
 	\label{fig:old_time_percentage}
 \end{figure*}

Below are descriptions of the AI prompt and several case studies of creating plots for the paper.
For a video walkthrough and the user study materials, see the replication package at \plotteryDotOrg~\suppMaterialsRef.

\subsection{AI Code-Generation Prompt}\label{appendix:prompt}

The AI prompt in TypeScript is:

\begin{notebooklisting}\begin{lstlisting}[keywords={},upquote=true]
const prompt =
`I am writing matplotlib code in the last cell of the following notebook:

\`\`\`
${notebook_code}
\`\`\`

Modify the code of Cell ${last_cell_no} to ${user_prompt}

Note: If I asked to create a function for a subplot and you need to pass in an ax argument, give that parameter a type annotation of mpl.axes._axes.Axes, but _only_ if there is a custom function for the subplot.

Note: If I asked above to add more subplots, use the following form:

\`\`\`
fig = plt.figure(...)
ax1 = fig.add_subplot(...)
ax2 = fig.add_subplot(...)
...
\`\`\`

but _only_ if I asked to add more subplots.

Return only the new code of Cell ${last_cell_no}`
\end{lstlisting}\end{notebooklisting}

\noindent
The \code{notebook\_code} is the code cells in the notebook (Markdown cells are excluded) joined into one long string, up to and including the current cell (later cells are not included). Each cell is prefixed with a comment like \code{\#\#\# Cell 1 \#\#\#} indicating the cell number.
The note about adding annotating the \code{ax} parameter was added after all the user studies and case studies were completed, it helps MyPy infer helpful types within the function.


\subsection{Case Studies}\label{sec:case-studies}
The user study tasks replicated existing plots. Here we present three brief case studies of using \sys to make real world plots \emph{from scratch}: namely, the plots in drafts of the paper. We focused on, but did not limit ourselves to, GUI interactions.

\mypara{Case Study 1 - Grouped Bar Charts}
We used \sys to create the grouped bar charts shown in \autoref{fig:old_interaction_counts} and \autoref{fig:old_time_percentage}, similar to the ``Group bar chart with labels'' example from the \mpl documentation.\footnote{\label{mpl-grouped-bars}\url{https://matplotlib.org/stable/gallery/lines_bars_and_markers/barchart.html}}
Grouped bar charts are difficult to implement in \mpl---there's no built-in function for it so the user must programmatically position and size each bar series manually.
In \sys, this can be done graphically.
The authoring is similar to the overview example in \autoref{sec:overview-example}, with few notable differences:

\begin{enumerate}[leftmargin=*]
  \item We create three overlapping bar series directly manipulate them on the plot to not only resize their widths but also translate them to be next to each other. The translation requires that their $x$ positions first be the the NumPy array \code{np.arange(len(}\hspace{0pt}\code{participants))} so that adding a single number during the mouse-drag (see \autoref{sec:implementation}) is a valid operation, because adding a number to an ordinary Python list is an error.
  \item We similarly position the legend simply by dragging it to the desired location.
  \item We abstracted the final plot into a function (possible via AI or code edits).
\end{enumerate}

\emph{In Summary: \sys allows fine-tune placement of on-plot elements by directly dragging and resizing, while the code facilitates later reusability through traditional coding abstractions, namely functions. This plot was particularly suited to \sys because it involved common \mpl functions with no data transformation.}

\begin{figure}[h]
\centering
\includegraphics[width=0.7\columnwidth]{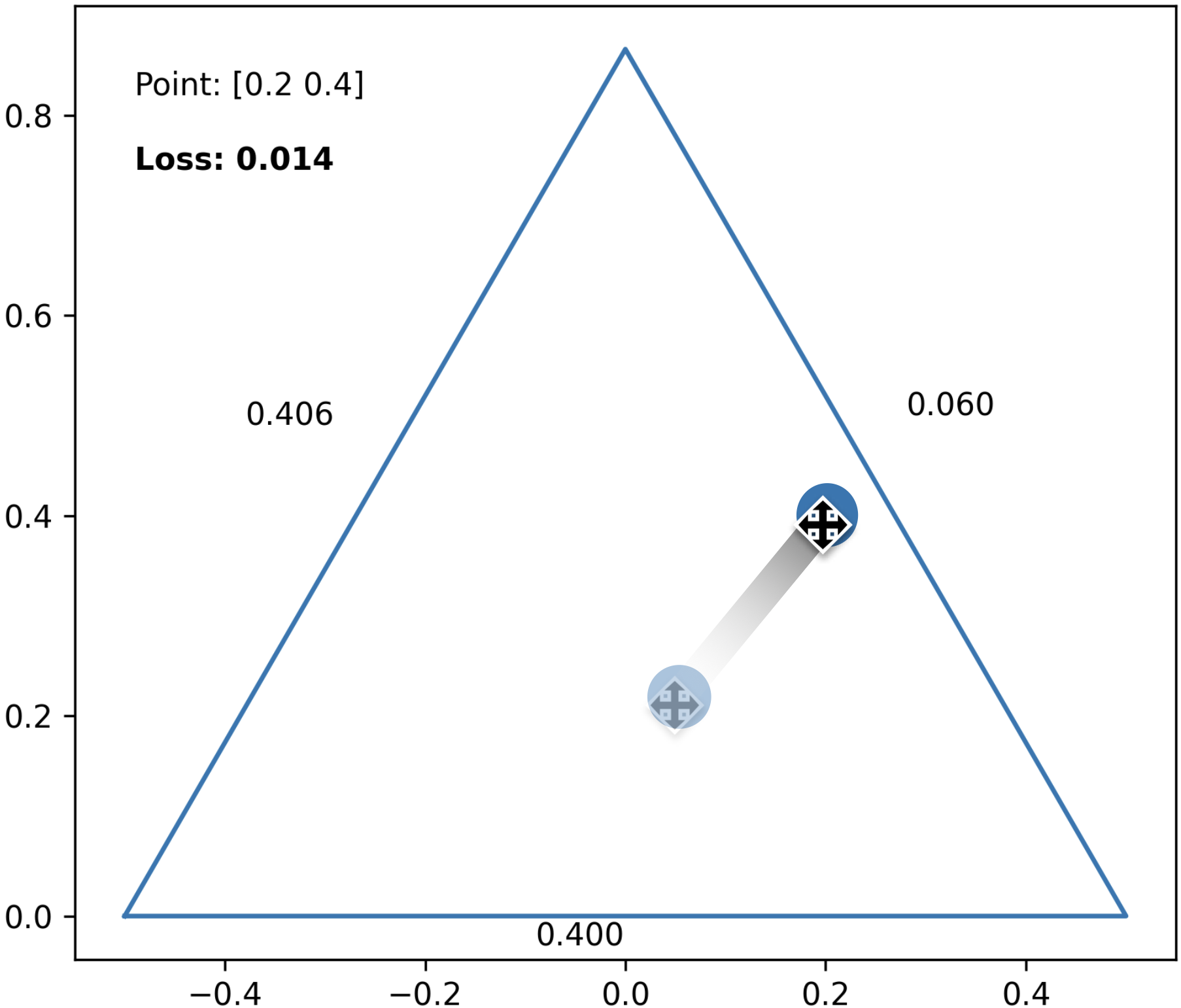}
\caption{An ad-hoc interactive plot for checking the calculation of the distance to the lines and the loss from the correct location.}
\label{fig:triangle_intermediate}
\Description{A triangle with numbers next to each edge.
A move cursor is dragging a dot around.
The numbers show the correct distance from the dot to each edge.
Text in the upper left says "Loss: 0.014"
}
\end{figure}

\mypara{Case Study 2 - Ternary Plot}
\sys also assisted us in creating the triangle plots in Figures~\ref{fig:old_interaction_counts} and \ref{fig:old_time_percentage}.
Again, this kind of plot is not built in to \mpl, so it requires a lot of manual placement.
The code for computing the $xy$ coordinates for the data points on the triangle was written manually.
We used \sys to check this code for correctness by constructing an ad-hoc interactive plot, shown in \autoref{fig:triangle_intermediate}.
We set up this plot so we could drag the dot around the plot and see live updates of the distance to the edges, shown in text next to each edge, as well as a ``Loss'' which should be 0 when the dot is at the right location.
%
Once satisfied these computations were correct—there were mistakes on the first try—we computed the $xy$ coordinates for all participants and created the plot.

\begin{figure}[h]
\centering
\includegraphics[width=0.5\columnwidth]{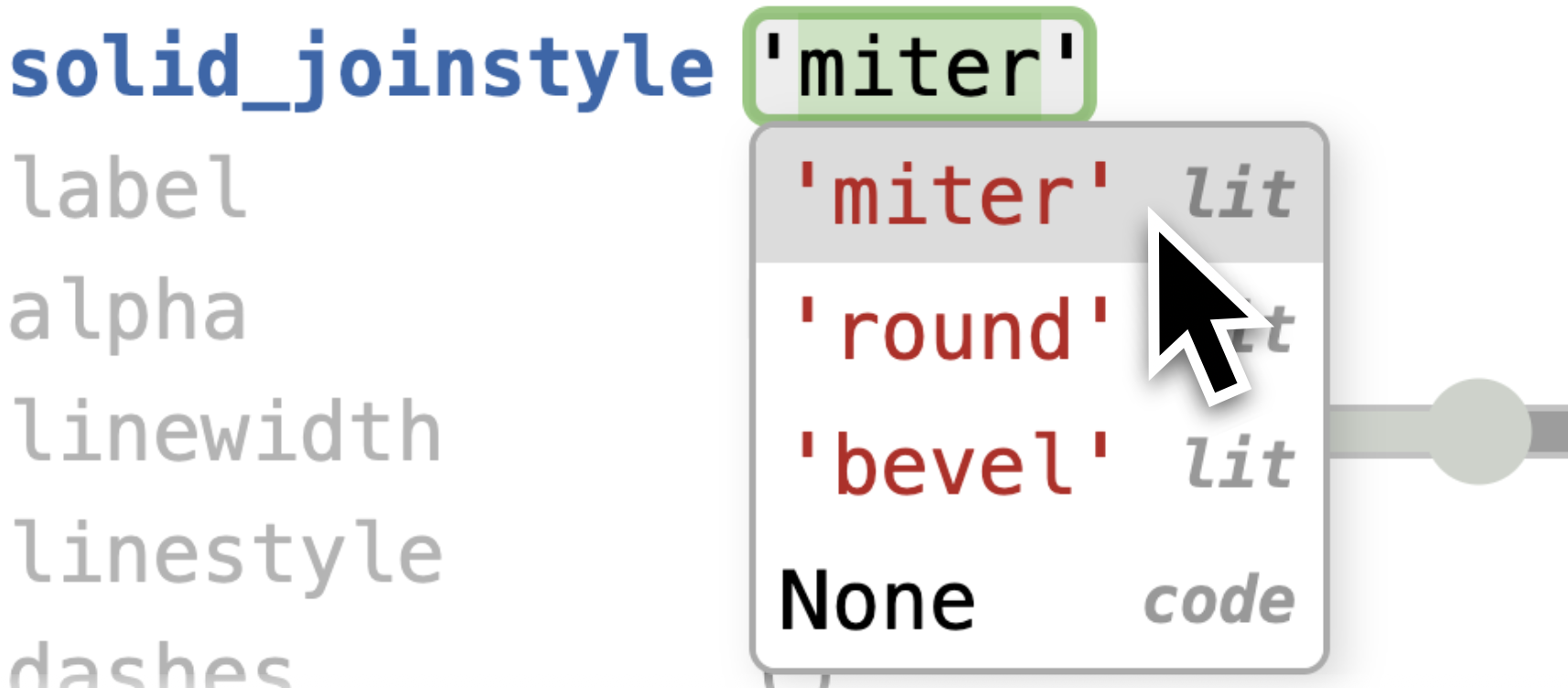}
\caption{Type-derived suggestions for \code{solid\_joinstyle}.}
\label{fig:case_study_solid_joinstyle}
\Description{The suggestions dropdown for solid_joinstype is open, with four suggestions: 'miter' 'round' 'bevel' and None. The cursor is over 'miter'}
\end{figure}

The initial plot was created without consulting the \mpl documentation.
For example, the default triangle corners were slightly rounded \raisebox{-0.3em}{\includegraphics[height=1.3em]{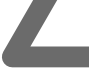}\Description{a round corner}} instead of sharp.
Rather than searching online, we toggled a few properties to find that \code{solid\_joinstyle} changed the corners, and we didn't need to look up the valid values because the API's type annotations produced the suggestions shown in \autoref{fig:case_study_solid_joinstyle}. The live preview instantly revealed that \code{'miter'} produced sharp corners \raisebox{-0.3em}{\includegraphics[height=1.3em]{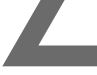}\Description{a sharp corner}}.
The Properties panel also helped style the text, but we had to repeat the edits three times for the labels around the triangles—first class support for linking properties via a shared variable would have helped.
We used AI to lock the aspect ratio of the $x$ and $y$ scales so the triangle was correctly equilateral, and with manual code edits we abstracted the plot into a function and extracted \code{border\_color} into a shared variable.

After adding more participants to our study, we revisited the plot to make the participant numbers more legible. Most notably, the AI helped us add the \code{bbox} argument to \code{ax.text}, which we might never have found by searching through documentation. Unfortunately, \sys's Properties panel could not adjust details of the \code{bbox} because it is a yet-unsupported \code{dict} object. Other GUI operations continued to work, for example when some numbers overlapped the existing text annotations (``No GUI'' on the code-AI edge) we were immediately able to drag the text annotation out of the way via the GUI.

The final versions shown in \autoref{fig:interaction_counts} and \autoref{fig:time_percentage} in the main paper were produced by a final round of re-styling. This was done without AI, mostly by GUI tweaking of visual properties and dragging items around. To get white outlines around the circle marks, code edits were used to duplicate the scatterplot and make a larger white version underneath. Code edits were also used to work out the math for the "knee" corner in the dotted "Mean" line on two of the plots. This offset amount for the corner was added as another parameter to the whole ternary plot drawing function, so the final amount of offset was tweaked using the GUI by selecting the appropriate drawing function call and dragging the numeric slider in the properties panel for the offset.


\emph{In Summary: \sys helped us create the ternary plots through code, GUI, and AI. The code was used to compute the data, and the GUI helped us verify our computations.
The GUI also allowed us to discover settings for corner shape and visual stylings, all in a live way, without the context-switching of having to look at documentation or query an external AI.
Still, the GUI supported only a subset of \mpl, inhibiting GUI explorations when the code was not in the supported subset.}

\begin{figure*}[t]
\centering
\includegraphics[width=0.7\textwidth]{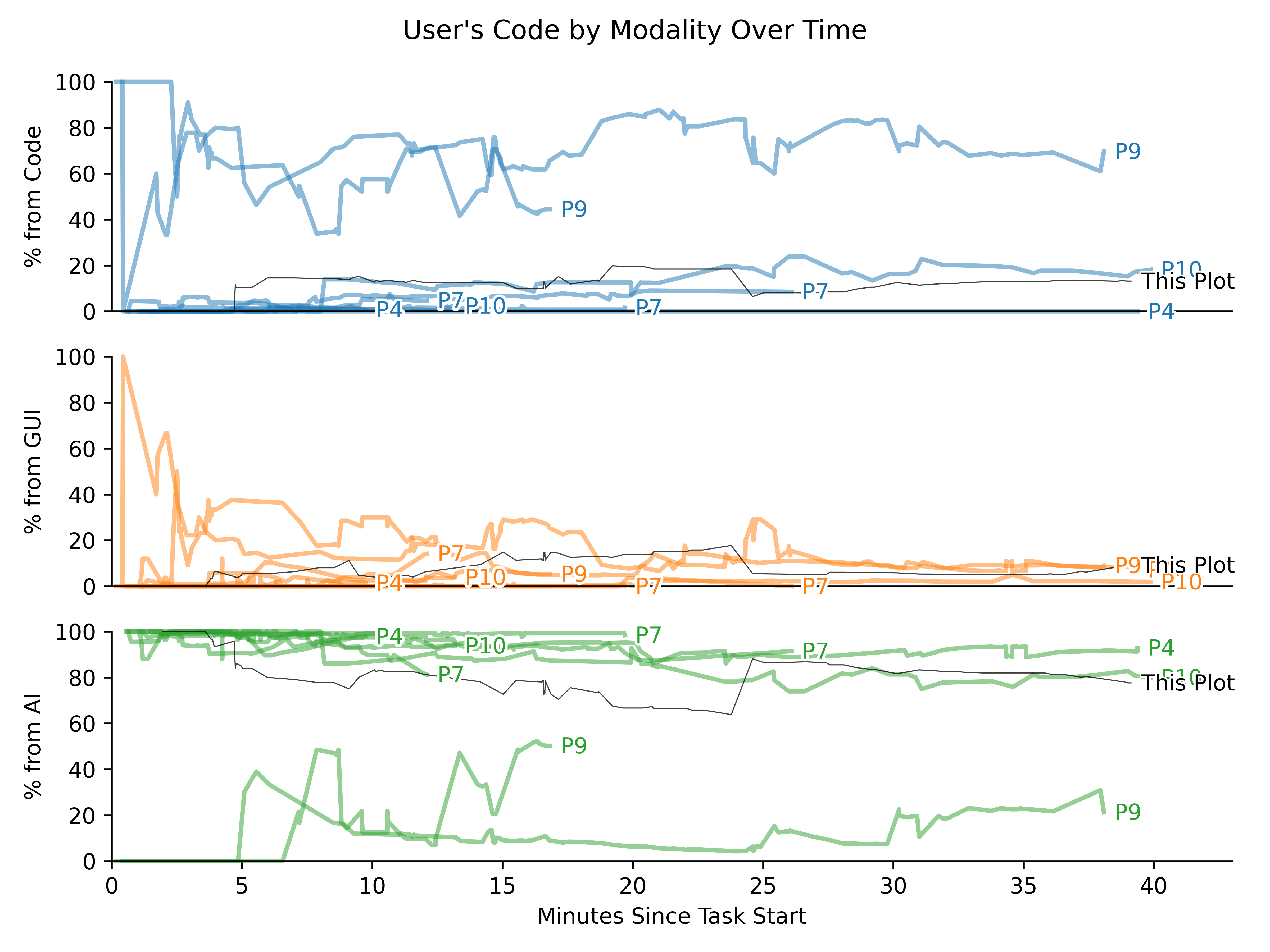}
\caption{The provenance of user's code over time for all \studytwo tasks. That is, if we stopped at time $t$, what proportions of the user's current cell code as seen at time $t$ came from each modality? Additionally, ``This Plot'' is the initial authoring of this plot by the first author. Most users started with AI and most code comes from AI, but all modalities contribute to authoring.}
\label{fig:code_by_modality_over_time}
\Description{Plot is titled "User's Code by Modality over Time"
3 subplots arranged vertically share an x axis of "Minutes Since Task Start" with a domain from 0 to 40.
The three subplot y axes are, respectively, "\% from Code", "\% from GUI", and "\% from Code", each with a range from 0 to 100.
Each subplot has 10 lines plotted, 9 are labeled with either P4, P7, P9, or P10, 1 is labeled "This Plot".
For the code subplot, the lines start at 0 and stay below 20\% except for P9's two tasks which quickly jump up and wobble between 40-80\%.
For the GUI subplot, the lines stay below 20\% except for one task of P9 which starts at 100\% and then drops to 20\% at 5 mins. P9's other task rises to 30\% at 5 mins and then wobbles slowly downward. All lines are generally decreasing over time.
For the AI subplot, all lines but P9's start at 100\% and rarely dip below 80\%. "This Plot" dips down to 70\% between minutes 20-24 before rising back to 80\%. P9's lines wobble between 5\% and 50\%.
}
\end{figure*}

\mypara{Case Study 3 - Code Modality Over Time}
Using the log of the participants interactions in \studytwo, we estimated where each part of the user's code came from in \studytwo---from code, GUI, or AI—by deducing the tokens of the code changed by each interaction.
For each task, \autoref{fig:code_by_modality_over_time} plots the proportion of the tokens in the user's currently visible code (excluding symbols, whitespace, and starter code) that were created (or most recently changed) by each modality as time progresses through each task.
It answers the question: if we stopped at time $t$, what proportions of the user’s current cell code as seen at time $t$ came from each modality?
Although arguably a good way to interpret the raw data, it is somewhat confusing so the figure was removed from the main paper.

The first author instrumented themselves creating the initial version of this plot and added that record as the ``This Plot'' line.
The plot shows that \P{4}, \P{7}, \P{10}, and the first author started with AI and AI remained responsible for the majority of their surviving code throughout authoring. \P{9} generally preferred more code and GUI editing. Below we describe the creation of \autoref{fig:code_by_modality_over_time} in \sys.

Before starting to plot \autoref{fig:code_by_modality_over_time}, we pasted the first few lines of our input CSV file into a comment so the AI would see the structure of our input data. 
We then chose to start by building a single subplot first because \sys's GUI is less helpful when there are multiple subplots—you must change each subplot individually.
 
 We used AI to plot the lines and to add the text labels at the end of each line. We performed code edits so the labels rendered on top of the lines and so they only showed participant ID and not task number. The GUI helped us fine-tune the position of these labels: we set the horizontal and vertical alignment via the Properties panel and directly dragged the labels around on the plot to change their offset. Dragging one moved all of them simultaneously because they came from a single call in a for-loop.
 We asked the AI to add a stroke to the labels using the ``patheffects'' library, using this specific vocabulary in our prompt so it would get it right on the first try (it did). The $y$ axis label and title, initially added by the AI, were modified through the GUI.
We then explored many variations to see if it were possible to style the lines so they could be individually distinguished. This exploration used all three modalities: \emph{AI} to scaffold the start of each experiment, \emph{code} to modify the scaffold (\eg style per participant instead of per line), and \emph{GUI} to explore many parameter tweaks. Eventually we decided the story was best told using thicker lines with partial transparency, without trying to distinguish the lines individually.
 
We asked the AI to abstract the single subplot code into a function and call it three times. This prompt worked, although, because of some unusal bug as mentioned in \autoref{sec:eval}, live feedback became unavailable and the interface lost its fluid feel. Separately, plot elements inside the function were also no longer selectable in the GUI—we have since changed the system AI prompt to avoid this issue (the \code{ax} input to the function needed a type annotation for MyPy). With degraded GUI support, we tweaked colors and axes limits with code edits, turned off the right and top axes spines through AI, and adjusted some whole-figure parameters through the GUI  including the whole-figure title, the figure size, and the resolution (these were outside the function and still GUI-editable).

 We instrumented ourselves as we created the plot and added that interaction log to the plot as the ``This Plot'' line. That final addition and minor style tweaks afterward are not included in ``This Plot''. The GUI degradation happened around minute 25.

\emph{In Summary: \sys's three modalities were all useful for constructing \autoref{fig:code_by_modality_over_time}. AI helped add the main elements onto the plot, the GUI helped tweak them and explore styling options, and code was used to do what the GUI could not or was too hard to describe to the AI. We did, however, lose some of \sys's features when we strayed to a code structure that was not encountered in the user studies or prior examples. The degraded GUI and loss of live feedback highlighted the convenience they provide when available.}


\begin{figure}[h]
\includegraphics[width=\columnwidth]{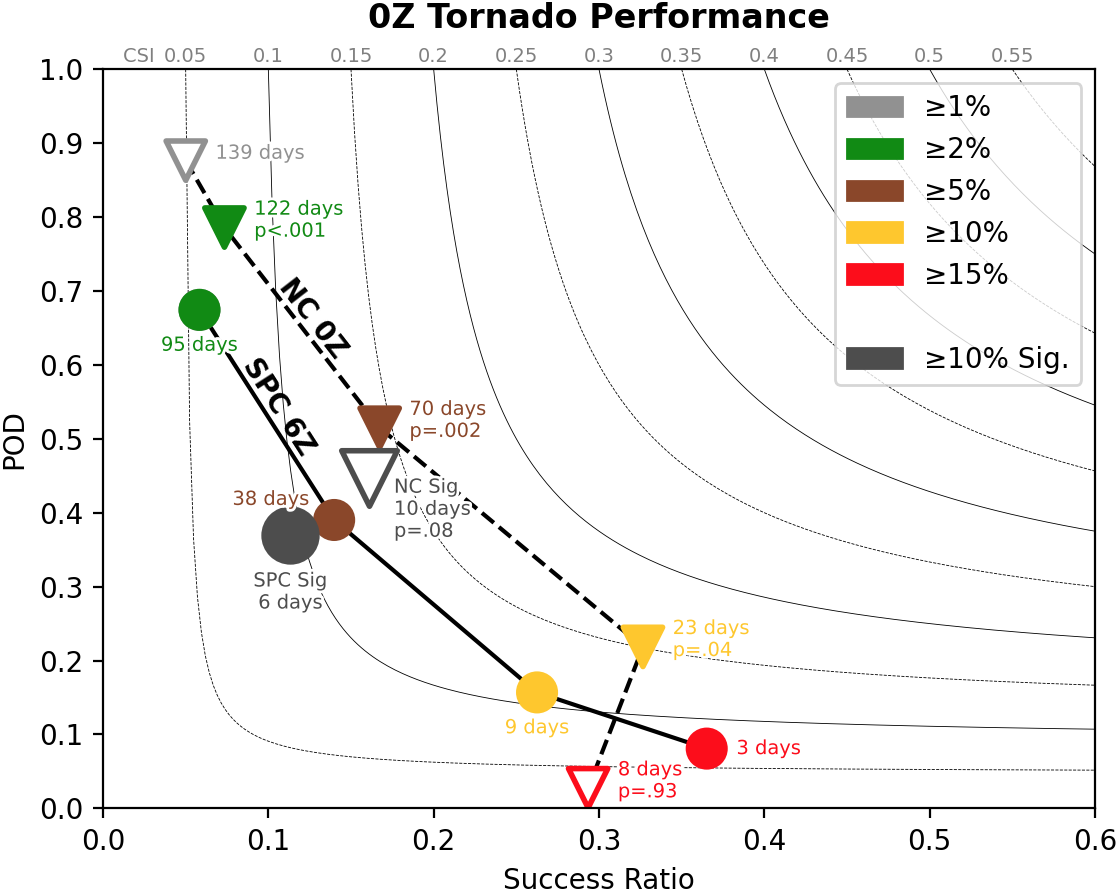}
\caption{A severe weather performance diagram.}
\label{fig:performance_diagram}
\end{figure}
\mypara{A Plot Too Complicated for \sys { - }Forecast Performance Plot}
\autoref{fig:performance_diagram} is a performance diagram\footnote{P. J. Roebber, ``Visualizing multiple measures of forecast quality,'' \emph{Weather and Forecasting}, vol. 24, no. 2, pp. 601 - 608, 2009. [Online]. Available: \url{https://journals.ametsoc.org/view/journals/wefo/24/2/2008waf2222159\_1.xml}}
from an unrelated manuscript in preparation by one of the authors. The plot depicts the probability of detection (POD, \ie recall) and the proportion of true positives (Success Ratio, \ie precision) for two tornado forecasting methods, ``SPC'' and ``NC'', at different forecasted probability thresholds (the different colors). Rightward and upward is better. NC's marks have a solid fill when they have a statistically significant difference from the SPC mark at the same threshold (same color).
 This plot predates \sys, but here we imagine how \sys might have—or, rather, might \emph{not} have—helped.
 The plotting code was crafted in a function because the manuscript draft itself has 12 such plots, and many more were created over the course of the research project. The function is 163 lines long and has 12 arguments, 12 for-loops and comprehensions, and 26 if-statements, allowing it to handle the many variations of the plot it must make. Only 40 of the 163 lines are \mpl calls.
 \sys might have helped make a ``first draft'' of the plot, but would become less useful as the functioned evolved with more special cases for variations.
 Nevertheless, a few improvements to \sys might prolong its usefulness had it been used to help author this plot:

 \begin{itemize}
   \item The function has a lot of code to build arrays of properties: \eg an array for the fill colors for a series of marks. An interface to craft simple list comprehensions might be helpful, \eg like Lyra~\cite{Lyra}'s simple spreadsheet interface.
   \item As mentioned in Case Study 2, for-loops to plot a text item or single mark are common. There are 6 such loops for this plot. A way to treat such simple loops as a first-class GUI entity would help—to \eg add filter predicates and simple local computations.
   \item The items in this legend had to be inserted manually with an impossible-to-remember incantation. GUI support for adding custom legend items would help as well.
 \end{itemize}

\subsection{On-Plot Direct Manipulation}

\begin{table}[h]
\caption{Drag Handle Argument Names \diffnote{Left-aligned text}}
\label{tab:drag_handle}
\centering
\small
\begin{tabular}{ll}
\toprule
\textbf{Drag Handle} & \textbf{Arg Name} \\
\midrule
Shape Body & \code{x} \emph{and/or} \code{y} \\
Top Edge  & \code{ymax} \emph{or} \code{top} \emph{or} \code{height} \\
Right Edge & \code{xmax} \emph{or} \code{right} \emph{or} \code{width} \\
Bottom Edge & \code{ymin} \emph{or} \code{bottom} \emph{or} \code{height} \\
Left Edge & \code{xmin} \emph{or} \code{left} \emph{or} \code{width} \\
Legend & \code{loc=(x, y)} \\
\bottomrule
\end{tabular}
\end{table}

\mypara{Notebook Extension}
\sys consists of a Jupyter Notebook extension, a Python backend, and a Typescript front-end that renders the GUI. The notebook extension installs a Jupyter hook to silently rewrite code before it is sent to the Python backend, replacing calls to \code{plt.show()} with 
a \sys-specific call that tells the Python backend about the variables in scope and the code in the notebook.
The call returns a Python object that renders the HTML and Javascript of \sys's interface (the object has a \code{\_repr\_html\_} method for Jupyter).

\end{document}